\documentclass[twocolumn]{aastex6}
\font\boldsym=cmmib10
\usepackage{graphicx}
\usepackage{epsfig}
\usepackage{natbib}
\usepackage{color}

\usepackage{hyperref} 
\hypersetup{colorlinks=true,citecolor=blue}
\usepackage{amsmath, amssymb, amsthm}
\newcommand     \mum    {\,\mu{\rm m}}  

\def \bea {\begin{eqnarray}}
\def \ena {\end{eqnarray}}

\def	\B	{{\rm B}}

\def 	\bE	{{\bf E}}

\def	\bJ	{{\bf J}}
\def	\bk	{{\bf k}}

\def    \bmu    {{\hbox{\boldsym\char'026}}}	

\def    \bomega {{\hbox{\boldsym\char'041}}}	

\def	\d	{{\rm d}}

\def	\eff	{{\rm eff}}
\def	\ed	{{\rm ed}}
\def	\ehat	{\hat{\bf e}}

\def	\gas	{\,{\rm gas}}

\def	\H	{{\rm H}}

\def	\LTE	{{\rm LTE}}

\def    \kB    {k_{\rm B}}

\def	\rot	{{\rm rot}}

\def	\tot 	{\rm {tot}}

\def	\Bar	{{\rm Bar}}

\def	\xhat		{\hat{\bf x}}
\def	\yhat		{\hat{\bf y}}
\def	\zhat		{\hat{\bf z}}
\def	\ahat		{\hat{\bf a}}
\def	\ehat		{\hat{\bf e}}

\def    \Bv     	{\bf  B}

\def    \kv     	{\bf  k}

\begin{document}
\title{Effect of alignment on polarized infrared emission from polycyclic aromatic hydrocarbons}
\author{Thiem Hoang}
\affil{Korea Astronomy and Space Science Institute, Daejeon 34055, Korea, email: thiemhoang@kasi.re.kr}
\affil{Korea University of Science and Technology, 217 Gajungro, Yuseong-gu, Daejeon, 34113, Korea}


\begin{abstract}
Polarized emission from polycyclic aromatic hydrocarbons (PAHs) potentially provides a new way to test basic physics of the alignment of ultrasmall grains. {In this paper, we present a new model of polarized PAH emission that takes into account the effect of PAH alignment with the magnetic field.} We first generate a large sample of the grain angular momentum $\bJ$ by simulating the alignment of PAHs due to resonance paramagnetic relaxation that accounts for various interaction processes. We then calculate the polarization level of PAH emission features, for the different phases of the ISM, including the cold neutral medium (CNM), reflection nebulae (RN), and photodissociation regions (PDRs). We find that a moderate degree of PAH alignment can significantly enhance the polarization degree of PAH emission compared to the previous results obtained with randomly oriented $\bJ$. In particular, we find that smallest, negatively charged PAHs in RN can be excited to slightly suprathermal rotation due to enhanced ion collisional excitation, resulting in an increase of the polarization with the ionization fraction. Our results suggest that RN is the most favorable environment to observe polarized PAH emission and to test alignment physics of nanoparticles. Finally, we present an explicit relationship between the polarization level of PAH emission and the degree of external alignment for the CNM and RN. The obtained relationship will be particularly useful for testing alignment physics of PAHs by future observations. 
 \end{abstract}

\section{Introduction}
Polycyclic aromatic hydrocarbons (PAHs) is an important dust component of the interstellar medium (ISM, \citealt{1984A&A...137L...5L}). PAH molecules are planar structures, consisting of carbon hexagonal rings and hydrogen atoms attached to their edge via valence bonds. Upon absorbing ultraviolet (UV) photons, PAHs reemit radiation in mid-infrared features, including 3.3, 6.2, 7.7, 8.6, 11.3, and 17 $\mu$m, due to vibrational transitions (see review by \citealt{2008ARA&A..46..289T}). Rapidly spinning PAHs also emit rotational radiation in microwaves via a new mechanism, so-called spinning dust (\citealt{1998ApJ...508..157D}; \citealt{Hoang:2010jy}). The latter is the most likely origin of anomalous microwave emission (AME) that contaminates Cosmic Microwave Background (CMB) radiation  (\citealt{Kogut:1996p5293}; \citealt{Leitch:1997p7359}). 

CMB experiments aiming to detect primordial gravitational waves through B-mode polarization face great challenges from polarized Galactic foregrounds (\citealt{Ade:2015ee}; \citealt{2016A&A...586A.133P}), including thermal dust emission and anomalous microwave emission (AME). Modern understanding shows that the AME is most likely produced by rapidly spinning nanoparticles, including PAHs, silicate (\citealt{2016ApJ...824...18H}; \citealt{Hensley:2016uf}) and iron nanoparticles \citep{2016ApJ...821...91H}. \footnote{Throughout this paper, dust grains refer to grains above 10 nm (or $100$\AA), while nanoparticles refer to ultrasmall grains smaller than ~10 nm.} The polarization level of these dust emission components depends on the alignment of interstellar grains and nanoparticles with the magnetic field. As a result, a quantitative description of grain alignment is required for accurate modeling of Galactic dust polarization. 

After more than 60 years since the discovery of starlight polarization by \cite{Hall:1949p5890} and \cite{Hiltner:1949p5856}, the longstanding problem of alignment of dust grains might be solved eventually (see latest reviews by \citealt{Andersson:2015bq} and \citealt{LAH15}). {The modern picture of grain alignment of {\it paramagnetic} grains can essentially be divided into two stages. First, the grain axis of maximum moment of inertia (e.g., short axis) is rapidly aligned with $\bJ$ due to Barnett relaxation (\citealt{1979ApJ...231..404P}).} Second, the angular momentum $\bJ$ is gradually aligned with the magnetic field by radiative torques (RATs) that are produced by interactions of anisotropic radiation field with helical grains (\citealt{1976Ap&SS..43..291D}; \citealt{1997ApJ...480..633D}; \citealt{2007MNRAS.378..910L}; \citealt{Hoang:2008gb}). The RAT alignment has become a leading mechanism to explain observational data. Fundamental predictions of the RAT alignment (\citealt{2009ApJ...695.1457H}; \citealt{2009ApJ...697.1316H}; \citealt{2014MNRAS.438..680H}; \citealt{2015MNRAS.448.1178H}) were observationally tested (\citealt{2011A&A...534A..19A}; \citealt{2013ApJ...775...84A}; \citealt{2014A&A...569L...1A}; \citealt{2015ApJ...812L...7V}). Very recently, \cite{2016ApJ...831..159H} proposed a unified model of grain alignment in which the joint action of radiative torques and enhanced magnetic relaxation by iron inclusions can produce perfect grain alignment. The unified model can successfully explain the high polarization level observed by {\it Planck} satellite and several observational puzzles, and allows self-consistent modeling of polarized dust emission. 

The basic physics of alignment of PAHs and ultrasmall grains \footnote{Hereafter, ultrasmall grains and nanoparticles are used interchangeably.} is still poorly understood. \cite{2000ApJ...536L..15L} first suggest that rapidly spinning nanoparticles can be weakly aligned by {\it resonance paramagnetic relaxation}, a modified version of Davis-Greenstein mechanism \citep{1951ApJ...114..206D} that works in rapidly spinning tiny grains. Numerical calculations in \cite{2014ApJ...790....6H} found that resonance relaxation can enable PAHs to be aligned up to a degree of $\sim10$ percent. However, \cite{2016ApJ...831...59D} pointed out that quantization effect may significantly suppress the alignment of nanoparticles because the vibrational energy levels are too broad compared to the intrinsic broadening width, freezing the energy transfer from rotational system to vibrational system. Meanwhile, \cite{2016MNRAS.457.1626P} suggested that {\it Faraday rotation braking} may enhance alignment of diamagnetic molecules (e.g., PAHs) due to direct transfer of rotational energy into heat via molecular vibrations. \footnote{The Faraday braking mechanism relies on the fact that a charge fixed to the rotating grain is moving with a velocity relative the grain center of mass.  In the presence of a magnetic field, the moving charge experiences Lorentz force that perturbs the charge motion. The coupling of the molecular vibration and motion induced by the Lorentz force is suggested to dissipate the rotation energy.} In this regard, polarization of PAH emission is a valuable way to test very physics of alignment of nanoparticles.

The polarization of starlight as well as of thermal dust emission can be modeled with a single alignment parameter, so-called Raleigh reduction factor (\citealt{Greenberg:1968p6020}), that describes an average degree of alignment of grain axis with the magnetic field. The polarization of PAH emission features, however, depends both on the internal alignment of the grain axis of major inertia $\ahat_{1}$ with $\bJ$ and the external alignment of $\bJ$ with the magnetic field $\Bv$. Modeling polarized PAH emission is thus complicated because PAH emission only occurs during a short time interval following UV photon absorption, {while the dynamical (e.g., rotational damping and grain alignment) timescales are much longer.} Therefore, to test the physics of alignment of nanoparticles, it is necessary to have a model that contains a direct link between the polarization level and grain alignment degrees. 

The first model of polarized PAH emission is presented by \cite{1988prco.book..769L} (hereafter L88) where the author noticed that internal alignment can produce polarized emission when PAHs being illuminated anisotropically by UV photons. For the 11.3 $\mu$m feature (out-of-plane C-H bending mode), the typical polarization level is estimated to be $\sim-2.1\%$ (L88), and the polarization direction is along the illumination direction (i.e., the central star-PAH molecule direction). For the 3.3$\mum$ feature (in-plane C-H stretching mode), the typical polarization is estimated to be $\sim 0.9\%$, with the polarization direction perpendicular to the illumination direction. {Here the positive and negative polarization corresponds to the polarization vector perpendicular and parallel to the illumination direction, respectively.} \cite{1988A&A...196..252S} (hereafter SDL88) observed the polarization of two PAH features and found the upper limit of $1\%$ for 3.3$\mum$ and $-3\%$ for 11.3$\mum$. \cite{Sironi:2009p5558} (hereafter SD09) revisited the L88's model by considering realistic rotational dynamics of PAHs. For a typical PAH molecule with $N_{\rm C}=200$ carbon atoms (radius of $a\sim 7$\AA), SD09 estimated the polarization fraction of $\sim 0.06\%$ for 3.3$\mum$ and $\sim -0.53\%$ for 11.3$\mum$, for the conditions of Orion Bar. 

We note that both L88 and SD09 assumed randomly oriented grain angular momentum in the space. Such an assumption underestimates the polarization level of PAH emission as mentioned in SD09 because PAHs are expected to be partially aligned due to paramagnetic resonance relaxation (\citealt{2000ApJ...536L..15L}). Indeed, \cite{2014ApJ...790....6H} found that resonance relaxation can enable PAHs in the CNM to be aligned with the magnetic field at a degree of external alignment $Q_{J}$ (see Appendix \ref{apdx:alignment} for definition) up to few percent for the typical dust temperature $T_{d}=60$ K, but it can increase to $10\%$ if PAHs can cool to a lower temperature of $20$ K between UV absorption events. This external alignment should produce higher polarization level than the case of random orientations of $\bJ$ in the space.

In the present paper, we will compute the polarization of PAH emission features by incorporating the effect of the partial alignment of $\bJ$ with $\Bv$ due to resonance paramagnetic relaxation mechanism. Moreover, we will employ latest progress in grain rotational dynamics (e.g., grain wobbling, anisotropic damping and excitation by IR emission) achieved in our previous works (\citealt{{Hoang:2010jy},{2011ApJ...741...87H}}; henceforth HDL10, HLD11). We aim to find a direct link between the polarization level of PAH emission with the degree of alignment with the magnetic field. This will pave the way for using polarized PAH emission to test basic physics of alignment of nanoparticles and for tracing magnetic fields using mid-IR PAH emission.

The structure of our paper is as follows. We first discuss relevant physics of PAHs and alignment mechanisms in Section \ref{sec:PAHalign}. In Section \ref{sec:PAHpol}, we describe the coordinate systems, numerical methods for calculations of polarization degree by including partial alignment of the angular momentum and the magnetic field. In Section \ref{sec:numpol}, we present our numerical results computed for the different environmental conditions. We discuss the implications of our obtained results in Section \ref{sec:discus}. A summary is presented in Section \ref{sec:sum}.

\section{Physics of PAHs and Grain Alignment}\label{sec:PAHalign}
\subsection{Magnetic properties of PAHs}
{ Ideal PAHs are expected to have rather low paramagnetic susceptibility due to H nuclear spin \citep{Jones:1967p2924}. However, astrophysical PAHs are likely magnetized thanks to the presence of free radicals, paramagnetic carbon rings, or adsorption of ions (see \citealt{2000ApJ...536L..15L}). 

We note that, during the last decade, significant progress has been made in research on magnetism of graphene, which provides insight into the magnetism of PAHs. For instance, \cite{2007PhRvB..75l5408Y} suggested that graphene can be magnetized by defects in carbon rings and adsorption of hydrogen atoms to the surface. The vacancy of carbon atoms from the carbon rings creates unpaired electrons, giving rise to the magnetization of graphene \citep{2007PhRvB..75l5408Y}. Also, the adsorption of a hydrogen atom induces magnetic ordering (see \citealt{2004PhRvL..93r7202L}), which is detected in a recent experiment (\citealt{2016Sci...352..437G}). In the ISM, the defects of PAHs can be triggered by bombardment of cosmic rays. All together, astrophysical PAHs are likely paramagnetic. Let $f_{p}$ be the fraction $f_{p}$ of paramagnetic atoms in the grain, and we take $f_{p}=0.01$ for calculations of PAH magnetic susceptibility, as in previous works (see \citealt{2014ApJ...790....6H} for details).}

\subsection{Barnett effect, Internal Relaxation and Internal Alignment}
\cite{Barnett:1915p6353} first pointed out that a rotating paramagnetic body can get magnetized with the instantaneous magnetic moment along the grain angular velocity $\bomega$. Later, \cite{1976Ap&SS..43..257D} introduced the magnetization via the Barnett effect for dust grains and considered its consequence on grain alignment. 

\cite{1979ApJ...231..404P} realized that the precession of $\bomega$ coupled to $\bmu_{\Bar}$ around the grain symmetry axis $\ahat_{1}$ produces a rotating magnetization component within the grain body coordinates. As a result, the grain rotational energy is gradually dissipated into heat until $\bomega$ becomes aligned with $\ahat_{1}$-- an effect that Purcell termed "Barnett relaxation". \cite{1999ApJ...520L..67L} revisited the problem by taking into account both spin-lattice and spin-spin relaxation (see \citealt{Morrish:1980}). Another internal relaxation process discussed in \cite{1979ApJ...231..404P} is related to the imperfect elasticity of the grain material, which was expected to be important for grains of suprathermal rotation only (see e.g., \citealt{1997ApJ...484..230L}).

{Internal relaxation (i.e., Barnett, nuclear relaxation, and imperfect elasticity) enables the transfer of grain rotational energy to the vibrational system. Naturally, some vibrational energy can also be transferred to the rotational system \citep{Jones:1967p2924}. For an isolated grain, a small amount of energy gained from the vibrational modes can induce fluctuations of the rotational energy $E_{\rot}$ when the grain angular momentum $\bJ$ is conserved (\citealt{1994MNRAS.268..713L}). Over time, the fluctuations in $E_{\rot}$ establish a local thermal equilibrium (LTE). 

Consider a planar PAH molecule with the axis of maximum moment of inertia $\ahat_{1}$ and the two other principal axes $\ahat_{2}\ahat_{3}$ in the PAH plane. Let $I_{\|} $ and $I_{\perp}$ be the moments of inertia along $\ahat_{1}$ and $\ahat_{2}$, and $h_{a}=I_{\|}/I_{\perp}$. The rotational energy is $E_{\rot}=J^{2}\left[1+(h_{a}-1)\sin^{2}\theta\right]/2I_{\|}$ where $\theta$ is the angle between $\ahat_{1}$ and $\bJ$. The fluctuations of the grain axis relative to $\bJ$ can be described by the Boltzmann distribution \citep{1997ApJ...484..230L}:}
\bea
f_{\LTE}(\theta,J)= Z{\exp}\left(-\frac{J^{2}}{2I_{\|}k_{\B}T_{ia}}
\left[1+(h_{a}-1)\sin^{2}\theta\right]\right),\label{eq:fLTE}
\ena
where $Z$ is a normalization constant such that $\int_{0}^{\pi} f_{\LTE}(\theta,J)\sin\theta d\theta=1$, and $T_{ia}$ is the internal alignment temperature above which the vibrational-rotational energy exchange is still effective.

\subsection{UV photon absorption and IR emission}

After a UV photon absorption, the grain vibrational energy is instantaneously increased to some maximum value. After a short time, the grain vibrational energy and temperature $T_{\rm vib}$ is reduced by emitting IR photons. The final temperature $T_{\rm ir}$ is determined by the vibrational temperature when the most IR photons are emitted. SD09 estimated $T_{\rm ir}=800, 300, 200$ and $120$ K for the $\lambda=3.3, 7.7, 11.3$ and $17\mum$ emission features, respectively. During the IR emission, the internal alignment temperature is equal to $T_{\rm ir}$ because of efficient energy exchange.

The grain temperature prior the UV absorption, $T_{0}$, is the same as the vibrational temperature as long as the vibrational-rotational energy exchange is still effective. We should stress that an exact determination of $T_{0}$ is challenging because it requires a detailed treatment of vibrational-rotational exchange that takes into account electron spin system (Barnett relaxation), inelastic effect, and quantum effect (\citealt{2016ApJ...831...59D}), which is beyond the scope of this paper. Instead, we will adopt the results from previous works. For the diffuse medium (e.g., CNM), previous works (SD09; \citealt{2011ApJ...741...87H}) find that PAHs can cool down to $T_{0}\sim 65$K in $10^{3} $s, much shorter than the time between two successive UV absorption events. For intense radiation conditions such as RN and PDR, the value $T_{0}$ is expected to be higher. 

\subsection{Magnetic Alignment of PAHs with the magnetic field by Resonance Paramagnetic Relaxation}
\cite{1951ApJ...114..206D} suggested that a paramagnetic grain rotating with angular velocity $\bomega$ in an external magnetic field $\Bv$ experiences paramagnetic relaxation due to the lag of magnetization, which dissipates the grain rotational energy into heat. This results in the gradual alignment of $\bomega$ and $\bJ$ with the magnetic field until the rotational energy is minimum.

For ultrasmall grains, such as PAHs, the classical Davis-Greenstein relaxation is suppressed because the rotation time is shorter than the electron-electron spin relaxation time $\tau_{2}$ (see \citealt{2014ApJ...790....6H}). Yet, such nanoparticles can be partially aligned by resonance paramagnetic relaxation that originates from the splitting of the rotational energy \citep{2000ApJ...536L..15L}. Numerical calculations in \cite{2014ApJ...790....6H} showed that PAHs can be aligned by resonance paramagnetic relaxation with the degree $Q_{J}\sim 0.05-0.15$ depending on $T_{0}$ and the magnetic field strength. Such a considerable degree of external alignment will produce anisotropy distribution of the angular momentum in the space, enhancing the polarization of PAH emission features. \cite{2016ApJ...821...91H} extended alignment calculations for iron nanoparticles and nanosilicates.

\section{Polarization of PAH emission features with grain alignment}\label{sec:PAHpol}
{In this section, we will describe in detail our new model of polarized PAH emission that treats the alignment of PAHs with the magnetic field. For a simple, intuitive model of polarized PAH emission, please refer to Appendix \ref{apdx:model} for details. }
\subsection{Coordinate systems}
\begin{figure*}\centering
\includegraphics[width=0.7\textwidth]{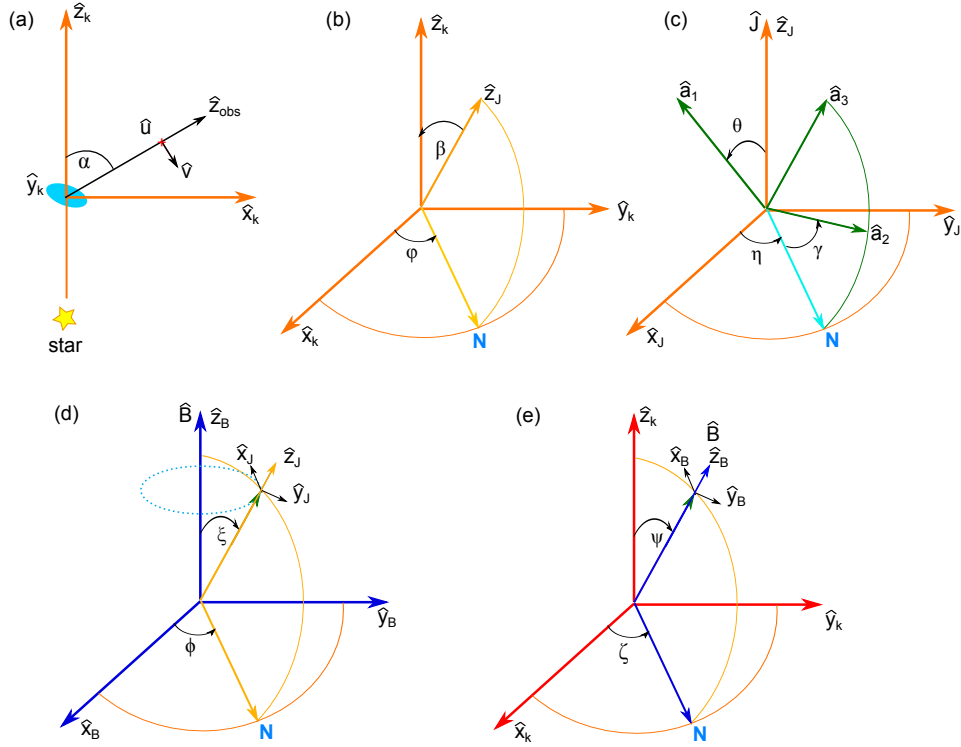}
\caption{Coordinate systems used in our calculations: (a) the star-molecule-observer system where $\zhat_{k}$ and $\zhat_{\rm obs}$ denote the illumination direction and the emission direction toward the observer; (b) orientation of the PAH angular momentum relative to the incident direction $\zhat_{k}$; (c) orientation of the PAH molecule in the reference frame defined by the angular momentum $\bJ$; (d) orientation of $\bJ$ in the magnetic field system; (e) orientation of the magnetic field in the system defined by the stellar incident radiation.}
\label{fig:RF}
\end{figure*}

Let consider a PAH molecule illuminated by UV radiation from a nearby star with the propagation direction $\zhat_{k}$. Assume that we are observing the PAH along the line of sight (LOS) $\zhat_{\rm obs}$ that makes an angle $\alpha$ with the incident radiation. Figure \ref{fig:RF} presents the various coordinate systems needed for calculations. The plane of the sky $\hat{\bf u}\hat{\bf v}$ is defined by unit vector $\hat{\bf v}$ in the plane $\zhat_{k}\zhat_{\rm obs}$ and $\hat{\bf u}$ perpendicular to the $\hat{\bf v}\zhat_{\rm obs}$ plane. Let $\xhat_{J}\yhat_{J}\zhat_{J}$ be unit vectors in which $\zhat_{J}$ is parallel to $\bJ$. Let $\xhat_{B}\yhat_{B}\zhat_{B}$ be unit vectors defined by the magnetic field such that $\zhat_{B}\|\Bv$.  

The angle between $\ahat_{1}$ and $\bJ$ is denoted by $\theta$ (panel (c)), and the angle between $\bJ$ and $\zhat_{k}$ is denoted by $\beta$  (panel (b)). The angular momentum $\bJ$ is then determined by the angles $\beta$ and $\varphi$ in the coordinate system defined by the incident radiation (panel (b)), and can also be described by the angles  $\xi$ and $\phi$ in the magnetic field coordinate system (panel (d)). The magnetic field direction is chosen to be fixed in the space and makes an angle $\psi$ with the radiation direction $\zhat_{k}$ and angle $\zeta$ with the plane $\xhat_{k}\zhat_{k}$ (panel (e)).

\subsection{PAH emission and Polarization}
Because IR emission occurs on a much shorter timescale compared to the time for the angular momentum variation (see L88), the flux of IR emission can be calculated for each momentum orientation $\bJ$.

Let $F_{u,v}^{\|,\perp}$ be the emission flux by a PAH molecule due to in-plane ($\|$) and out-of-plane oscillation ($\perp$) with the electric field $\bE$ in the $\hat{\bf u}$ and $\hat{\bf v}$ directions, respectively (see Figure \ref{fig:RF}(a)). Let $I_{u,v}$ be the total emission intensity from the PAH. 

The emission flux $F_{u,v}^{\|,\perp}$ depends on the orientation of the PAH plane with $\bJ$ and the distribution of $\bJ$ with the magnetic field, which are described by the distributions $f_{\rm LTE}(\theta, J)$ (Equation \ref{eq:fLTE}) and $f_{J}(\bJ)$, respectively. Thus, the total emission intensity is obtained by integrating over these distribution functions:
\bea
I_{\star,w}^{\|,\perp}(\alpha, \psi) &=& \int f_{J}(\bJ)d\bJ\int_{0}^{\pi}f_{0}(\theta_{0},J)d\theta_{0}\int_{0}^{\pi}f_{ir}(\theta, J)d\theta\nonumber\\
&&\times A_{\star}(\beta,\theta_{0})F_{w}^{\|,\perp}(\beta, \varphi,\theta,\alpha),\label{eq:Istar_uv}
\ena
where $w=u,v$, $A_{\star}$ is the cross-section of UV absorption, and $f_{0},f_{ir}$ are given by Equation (\ref{eq:fLTE}) with $T_{ia}=T_{0}$ and $T_{ir}$ (see L88, SD09). The explicit expressions of $A_{\star}$ and $F_{w}$ are given in Appendix \ref{apd:A}. The emission intensity $I_{\star,w}(\alpha, \psi)$ is only a function of $\alpha$ and $\psi$, which describes the angles of the illumination radiation relative to the LOS and the magnetic field (see Figure \ref{fig:RF}).

The polarization degree of PAH emission due to in-plane and out-of-plane oscillations is calculated as the following:
\bea
p^{\|,\perp}(\alpha, \psi)=\frac{I_{u}^{\|,\perp}(\alpha, \psi)-I_{v}^{\|,\perp}(\alpha, \psi)}{I_{u}^{\|,\perp}(\alpha, \psi)+I_{v}^{\|,\perp}(\alpha, \psi)},\label{eq:pol_a}
\ena
{where the positive and negative $p$ correspond to the polarization vector along the $\hat{\bf u}$- and $\hat{\bf v}$- direction, respectively (see Figure \ref{fig:RF}). }

The polarization by a population of PAHs with size distribution $dn/da$ is
\bea
p^{\|,\perp}(\alpha, \psi)=\frac{I_{u}^{\|,\perp}(\alpha, \psi)-I_{v}^{\|,\perp}(\alpha, \psi)}{I_{u}^{\|,\perp}(\alpha, \psi)+I_{v}^{\|,\perp}(\alpha, \psi)},\label{eq:pol}
\ena
where the intensity obtained by integrating over the grain size distribution:
\bea
I_{\star,w}^{\|,\perp}(\alpha,\psi)= \int da 4\pi a^{2}(dn/da)I_{\star,w}^{\|,\perp}(\alpha,\psi).\label{eq:Istar}
\ena

\subsection{Numerical integration}
Since the emission flux $F_{w}^{\|,\perp}$ is a function of $\beta,\varphi$ while the $\bJ$ orientation obtained from the Langevin equations is described by $(\xi,\phi)$ in the magnetic field frame, we first transform $(\xi,\phi)$ to $(\beta,\varphi)$ using the following equations (see Appendix \ref{apd:transform}):
\bea
\cos\beta &=& \cos\xi\cos\psi -\sin\xi\cos\phi\sin\psi,\label{eq:cos_beta}\\
\sin\beta\cos\varphi &=&\cos\xi\sin\psi\cos\zeta \nonumber\\
&& + \sin\xi\left(\cos\phi\cos\psi\cos\zeta -\sin\phi\sin\zeta \right),\label{eq:cos_varphi}
\ena
where $\psi$ and $\zeta$ are shown in Figure \ref{fig:RF}(panel (e)).

Next, we make use of ergodic approximation of the grain dynamical system to numerically compute the emission intensity given by Equation (\ref{eq:Istar_uv}). Basically, we can replace the ensemble average (i.e., over the angular distribution $f_{J}(\bJ)dJ_{x}dJ_{y}dJ_{z}$) by time average over all possible orientations and values of the grain angular momentum (ergodic theory). Thus, the emission intensity can be calculated as
\bea
I_{\star,w}^{\|,\perp}=\int f_{J}(\bJ)d\bJ\times\mathcal{I}_{w}^{\|,\perp} = \frac{1}{N}\sum_{\lbrace J,\beta,\varphi\rbrace_{i};i=1}^{i=N}\mathcal{I}_{w}^{\|,\perp}(J_{i},\beta_{i},\varphi_{i}),~~\label{eq:Iw_num}
\ena
where $\mathcal{I}$ denotes the entire term after $f_{J}(\bJ)d\bJ$ in Equation (\ref{eq:Istar_uv}).

The polarization degree is then calculated by Equation (\ref{eq:pol_a}). For the case in which the magnetic field lies in the plane of the sky, we have $\zeta=\pi/2$. When the magnetic field is directed along the radiation direction, $\xi\equiv \beta$ and $\phi\equiv \varphi$. In this case, the fast Larmor precession allows averaging over the azimuthal angle $\varphi$ of $\bJ$ around the illumination direction $\zhat$.

\subsection{Simulations of grain angular momentum}\label{sec:LE}
In this section, we briefly describe our numerical method to create a large sample of angular momentum from numerical simulations, $\lbrace \bJ\rbrace_{i}\equiv \lbrace J,\xi,\varphi \rbrace_{i}$.

Let $a$ is the effective size defined as the radius of an equivalent sphere of the same volume as the PAH molecule. We adopt the model of PAHs as in HDL10, where small PAHs of size $a\le 6$\AA~ are assumed to have disk-like shape of thickness $d=3.3$\AA. Larger PAHs are assumed to be spherical. Since the timescale for a change in the angular momentum is much longer than the IR emission time, we take the temperature $T_{0}$ for PAHs in calculations of $f(\bJ)$.

As in previous works (\citealt{1999MNRAS.305..615R}; \citealt{2014ApJ...790....6H}), to find $\lbrace\bJ\rbrace_{i}$, we solve the Langevin equations for the evolution of $\bJ$ in time in an inertial coordinate system denoted by unit vectors $\ehat_{1}\ehat_{2}\ehat_{3}$ where $\ehat_{1}$ is chosen to be parallel to $\Bv$. The Langevin equations read 
\bea
dJ_{i}=A_{i}dt+\sqrt{B_{ii}}dW_{i}\mbox{~for~} i=~1,~2,~3,\label{eq:dJ_dt}
\ena
where $dW_{i}$ are the random variables drawn from a normal distribution with zero mean and variance $\langle dW_{i}^{2}\rangle=dt$, and $A_{i}=\langle {\Delta J_{i}}/{\Delta t}\rangle$ and $B_{ii}=\langle \left({\Delta J_{i}}\right)^{2}/{\Delta t}\rangle$ are the drifting (damping) and diffusion coefficients defined in the $\ehat_{1}\ehat_{2}\ehat_{3}$ system. Detailed descriptions of the diffusion coefficients are presented in HDL10 (see also \citealt{1998ApJ...508..157D}).

It is convenient to write the Langevin equations in the dimensionless units of $J'\equiv J/I_{\|}\omega_{T}$ and $t'\equiv t/\tau_{\H,\|}$ where $\omega_{T}$ is the thermal angular velocity at gas temperature $T_{\gas}$, and $\tau_{\H,\|}$ is the rotational damping time along the symmetry axis (see HDL10). Thus, Equation (\ref{eq:dJ_dt}) becomes 
\bea
dJ'_{i}=A'_{i}dt'+\sqrt{B'_{ii}}dw'_{i} \mbox{~for~} i= 1,~2,~3,\label{eq:dJp_dt}
\ena
where $\langle dw_{i}^{'2}\rangle=dt'$ and
\bea
A'_{i}&=&-{J'_{i}}\left[\frac{1}{\tau'_{\gas,{\eff}}} +\delta_{m}(1-\delta_{1i})\right] -\frac{2}{3}\frac{J_{i}^{'3}}
{\tau'_{\ed,{\eff}}},\label{eq:Ai}~~~~\\
B'_{ii}&=&\frac{B_{ii}}{2I_{\|}\kB T_{\gas}}\tau_{\H,\|}+\frac{T_{\d}}{T_{\gas}}\delta_{\rm m}(1-\delta_{1i}),\label{eq:Bii}
\ena
where $T_{d}$ is the dust temperature (same as $T_{0}$). Above, $\delta_{m}=\tau_{\H,\|}/\tau_{m}$ with $\tau_{m}$ magnetic alignment timescale (see \citealt{2014ApJ...790....6H}), $\delta_{1i}=1$ for $i=1$ and $\delta_{1i}=0$ for $i\ne 1$, and
\bea
\tau'_{\gas,{\eff}}= \frac{\tau_{\gas,{\eff}}}{\tau_{\H,\|}},~\tau'_{\ed,{\eff}}&=&\frac{\tau_{\ed,{\eff}}}{\tau_{\H,\|}},~~~
\ena
where $\tau_{\gas,{\eff}}$ and $\tau_{\ed,{\eff}}$ are the effective damping times due to dust-gas interactions and electric dipole emission (see Eq. E4 in HDL10). 

To numerically solve the Langevin equations (\ref{eq:dJp_dt}), we follow the approach in \cite{2016ApJ...821...91H} (hereafter HL16) where a second-order integrator is applied. The angular momentum component $j_{i}\equiv J'_{i}$ at iterative step $n+1$ is evaluated as follows:
\bea
{j}_{i;n+1} &=& j_{i;n} - \gamma_{i}{j}_{i;n}h+\sqrt{h}\sigma_{ii}{\zeta}_{n}- \gamma_{i}\mathcal{A}_{i;n}-\gamma_{\ed}\mathcal{B}_{i;n},~~~
\ena
where $h$ is the timestep, $\gamma_{i}=1/\tau'_{\gas,{\eff}} +\delta_{m}(1-\delta_{zi})$, $\gamma_{\ed}=2/(3\tau'_{\ed,\eff})$, $\sigma_{ii}=\sqrt{B'_{ii}}$, and
\bea
\mathcal{A}_{i;n}&=& -\frac{h^{2}}{2} \gamma_{i} j_{i;n}+\sigma_{ii} h^{3/2}g(\zeta_{n},\eta_{n})-\gamma_{\ed}j_{i;n}^{3}\frac{h^{2}}{2},\\
\mathcal{B}_{i;n}&=&j_{i;n}^{3}h - 3\gamma_{i}j_{i;n}^{3}\frac{h^{2}}{2}-\frac{3j_{i;n}^{5}\gamma_{\ed}h^{2}}{2}+3j_{i;n}^{2}\sigma_{ii} h^{3/2}g(\zeta_{n},\eta_{n}),
\ena
with ${\eta}_{n}$ and ${\zeta}_{n}$ being independent Gaussian variables with zero mean and unit variance and $g(\zeta_{n},\eta_{n})= \zeta_{n}^{2} /2 + \eta_{n}^{2} /2\sqrt{3}$ (see Appendix C in HL16 for details).

The timestep $h$ is chosen by $h= 0.01\min[1/F_{\tot,\|}, 1/G_{\tot,\|}, \tau_{\rm ed,\|}/\tau_{\H,\|},1/\delta_{m}]$.  As usual, the initial grain angular momentum is assumed to have random orientation in the space and magnitude $J=I_{\|}\omega_{T}$ (i.e., $j=1$).

For PAHs, the timestep $h$ is determined essentially by two timescales $\tau_{\rm gas}$ and $\tau_{\ed}$. The difference between these two timescales can be huge, up to an order of $10^{4}$ for CNM. In this case, a tiny timestep $h$ is needed to achieve sufficient statistics, which requires a huge number of time steps and large computing time. In the following, a fixed integration time $T=100\tau_{\gas}$ is chosen, which ensures that $T$ is much larger than the longest dynamical timescale to provide good statistics of the degrees of grain alignment. Thus, we can generate up to $N\sim 10^{7}$ possible orientations of $\bJ$ in space, described by $\lbrace J,\xi,\phi\rbrace_{i}$. The grain rotational temperature can be calculated as $T_{\rot}=\langle J^{2}\rangle/k_{\B}I_{\|}$ where $\langle J^{2}\rangle =\sum_{i=1}^{N}J^{2}/N$. The degree of internal alignment and external alignment, $Q_{X}$ and $Q_{J}$, are calculated as described in Appendix \ref{apdx:alignment}.

\section{Numerical Results}\label{sec:numpol}
\subsection{Code Benchmark}
To benchmark the numerical integration code through ensemble averaging, we generate a sample of $N$ different orientations of $\bJ$ that follows the isotropic distribution in the space, assuming a constant value of $J$. For each orientation, we compute $\psi=2\pi u, \beta=\cos^{-1}(2v-1)$ with $u,v$ are the random variables drawn from the uniform distribution in the range of $(0,1)$. We first test the mean value of $\cos^{2}\beta$ obtained from the ensemble averaging against the correct value. We found that $\langle \cos^{2}\beta\rangle\equiv \sum_{i=1}^{N}\cos^{2}\beta_{i}/N \approx 1/3\equiv \int_{0}^{\pi}\cos^{2}\beta d\cos\beta/2$ for $N> 10^{4}$. Then, we compute the polarization of PAH emission using Equation (\ref{eq:pol}) for a sample of $N$ random orientations with a constant rotational temperature $T_{\rm rot}=T_{\gas}$ to compare with the analytical result derived for randomly oriented $\bJ$ from SD09. We found that an good agreement is achieved for $N\ge 10^{5}$. 

Finally, we compute the polarization of PAH emission using the simulated angular momentum data $\lbrace J, \xi, \phi\rbrace_{i}$ obtained from the Langevin equation simulations with the use of the coordinate transformations (Equations \ref{eq:cos_beta}-\ref{eq:cos_varphi}). We calculate the polarization degree by averaging over a sample $N=10^{6}-10^{7}$, where the initial step $N_{i}=10^{6}$ is long enough for the system to forget its initial conditions.

\subsection{Model setup}
We compute the polarization degree of PAH emission for the different physical conditions, including CNM, RN, PDR, and Orion Bar. Table \ref{tab:ISM} presents the typical physical parameters for these idealized environments where $n_{\H}$ is the hydrogen number density, $T_{\gas}$ and $T_{0}$ are gas and dust temperature, $\chi = u_{\rm rad}/u_{\rm ISRF}$ is the ratio of the radiation energy density $u_{\rm rad}$ to the radiation density of the average diffuse ISM in the solar neighborhood, u$_{\rm ISRF}$ (see \citealt{1983A&A...128..212M}), and n(H$_{2}$), n(H$^{+}$), and n(C$^{+}$) are the molecular hydrogen density, ion hydrogen density, and ionized C density, respectively. 

We consider several strong emission features, including in-plane modes, $3.3$ and $7.7\mum$, and out-of-plane modes, $11.3\mum$ and 17$\mum$. We assume that the incident radiation arrives at the right angle with the line of sight, i.e., $\alpha=90^{\circ}$, corresponding to the upper limit of the expected polarization. The magnetic field is parallel to the stellar radiation direction (i.e., $\psi=0^{\circ}$), and both {the magnetic field and stellar radiation directions} lie in the plane of the sky, unless stated otherwise.

Let $\gamma_{0}=T_{\rm rot}/T_{0}$ and $\gamma_{\rm ir}=T_{\rm rot}/T_{\rm ir}$ where $T_{\rm rot}$ is the rotational temperature. These parameters $\gamma_{0},\gamma_{\rm ir}$ determine the degree of internal alignment prior UV absorption and during the IR emission, respectively (SD09). In addition, we introduce a new parameter, namely, $\gamma_{\rm sup}=T_{\rm rot}/T_{\gas}$, describing the degree of external alignment of the angular momentum with the magnetic field. We determine $\gamma_{0},\gamma_{\rm ir},\gamma_{\rm sup}$ from simulations of Langevin equations.

\begin{table}
\caption{Typical physical parameters of the selected interstellar phases}\label{tab:ISM}
\begin{tabular}{l l l l l l} \hline\hline\\
\multicolumn{1}{c}{\it Parameters} & {CNM} & RN & PDR & Orion Bar\\[1mm]
\hline\\
$n_{\rm H}$~(cm$^{-3}$) &30 &$10^{3}$ & $10^{5}$ & $7\times 10^{4}$\\[1mm]
$T_{\rm gas}$~(K)& 100  &100 & $10^{3}$ & $10^{3}$\\[1mm]
$T_{\rm 0}$~(K)& 20  &40 &80 & 150\\[1mm]
$\chi$ &1 &$10^{3}$ & $3\times 10^{4}$& $3\times 10^{4}$\\[1mm]
$x_{\rm H}=n(\H^{+})/n_{\H}$ &0.0012  &$10^{-3}$ & $10^{-4}$ &$10^{-4}$\\[1mm]
$x_{\rm C}=n(\rm C^{+})/n_{\H}$ &$3\times 10^{-4}$  &$2\times 10^{-4}$ &$2\times 10^{-4}$&$2\times 10^{-4}$\\[1mm]
$y=2n({\rm H}_{2})/n_{\rm H}$&{$0.$} &0.5 &0.1 & 0.1 \\[1mm]
$B~(\mu G)$&{$10$} & 100 & 100 & 100\\[1mm]
\\[1mm]
\hline\hline\\
\end{tabular}
\end{table}

\subsection{Cold Neutral Medium}
We first present the numerical results of polarized PAH emission from CNM. Three different grain temperatures before UV absorption, $T_{0}=20, 40$K and 60K, are assumed.

\begin{figure*}
\centering
\includegraphics[width=0.4\textwidth]{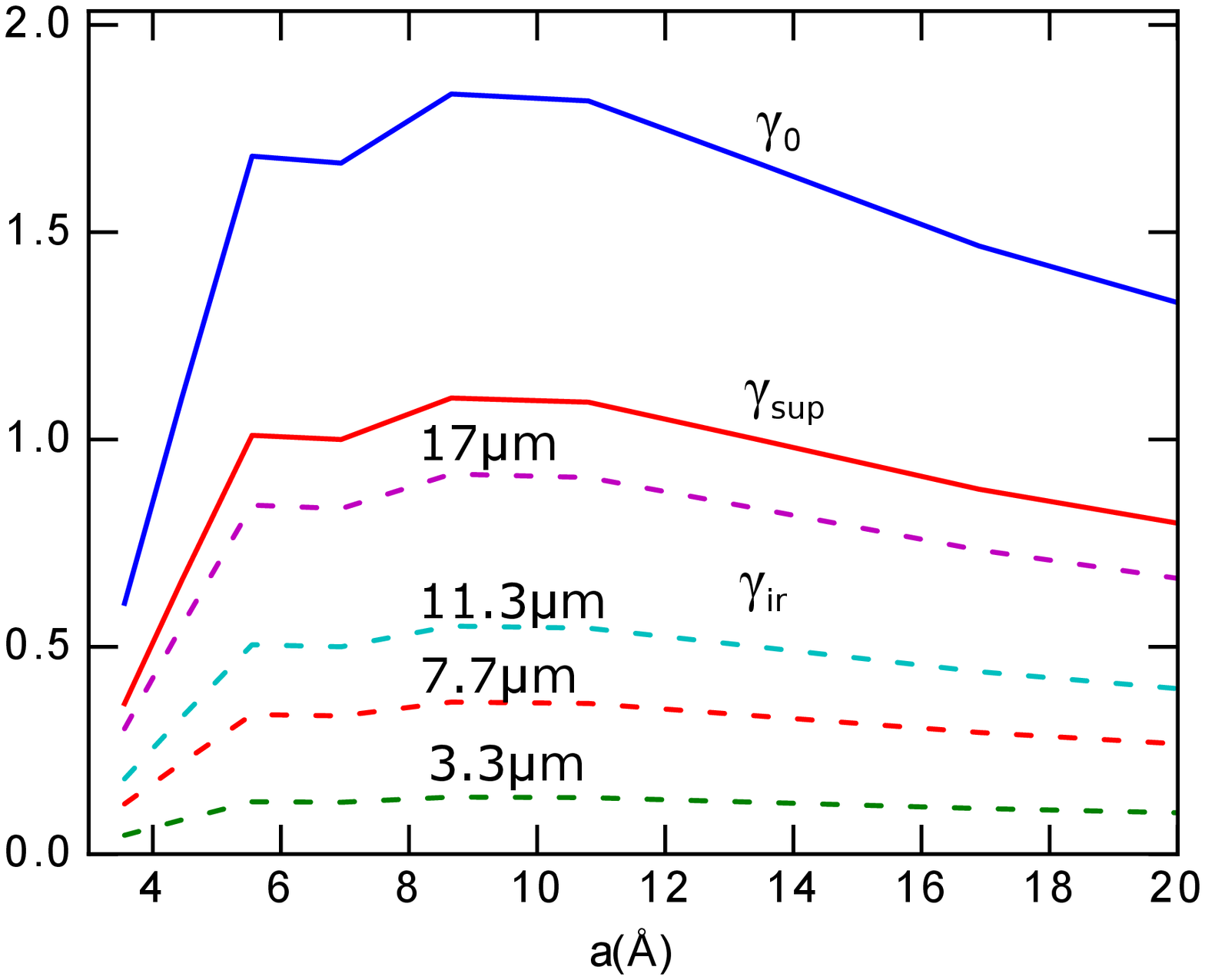}
\includegraphics[width=0.4\textwidth]{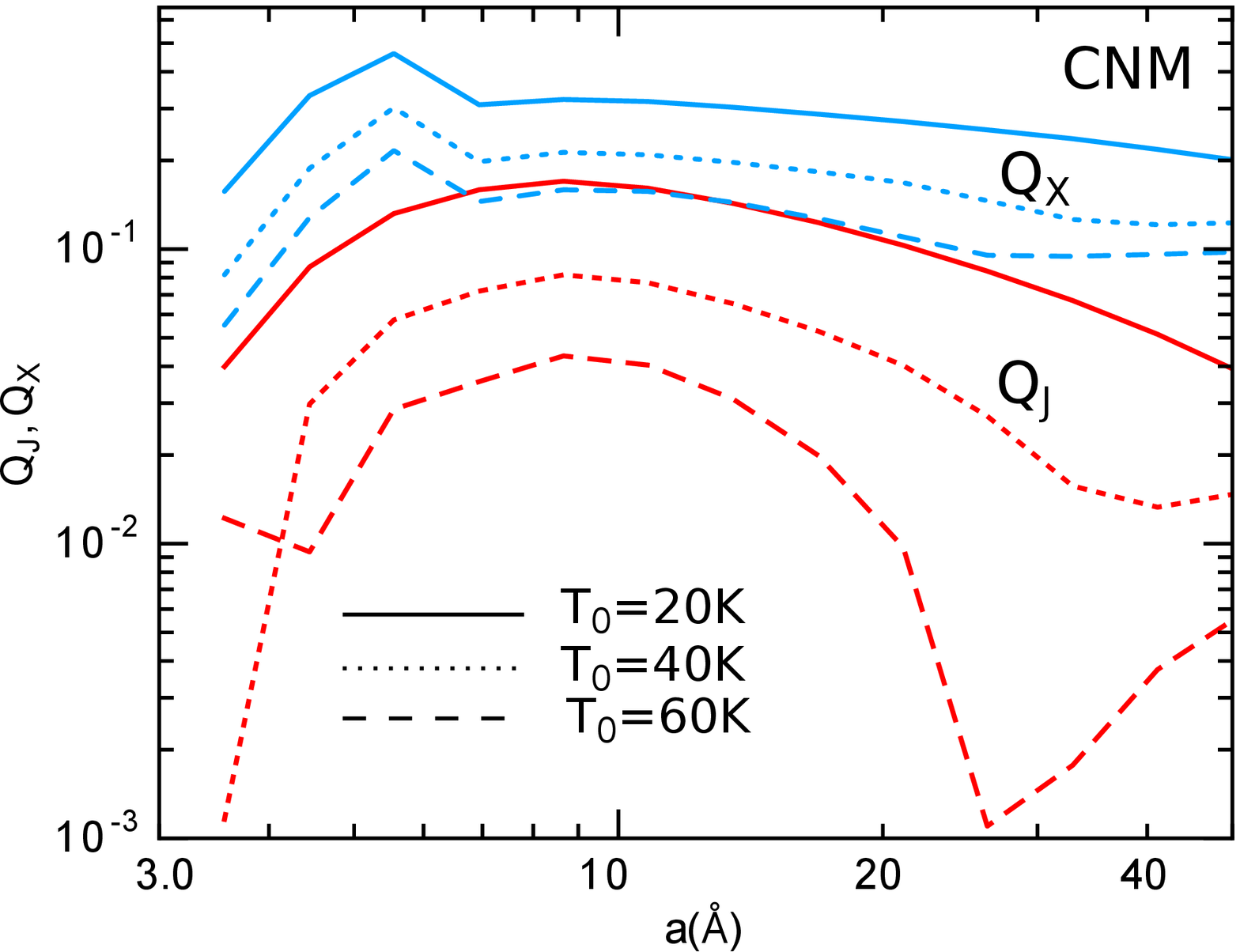}
\caption{Left panel: $\gamma_{\rm sup}$, $\gamma_{0}$ and $\gamma_{\rm ir}$ vs. grain size $a$ at $T_{0}=60$ K. Right panel: the degree of internal alignment ($Q_{X}$) and external alignment ($Q_{J}$) computed for the different temperatures $T_{0}$, where $Q_{J}$ peaks at $a\sim 8.5$\AA.}
\label{fig:QXQJ_CNM}
\end{figure*}

Figure \ref{fig:QXQJ_CNM} shows the values of $\gamma_{\rm sup},\gamma_{0},\gamma_{\rm ir}$ for the typical temperature $T_{0}=60$ K (left) and the degree of internal alignment ($Q_{X}$) and external alignment ($Q_{J}$) (right panel) for the three considered temperatures. The value of $\gamma_{sup,0,ir}$ declines rapidly from its peak for $a<8$\AA~ due to the decrease of $T_{\rm rot}$ as a result of increased rotational damping by electric dipole emission (\citealt{1998ApJ...508..157D}; \citealt{2014ApJ...790....6H}). Moreover, the value of $Q_{J}$ is below $0.05$ for $T_{0}=60$ K and increases to $Q_{J}\sim 0.15$ when the temperature is decreased to $T_{0}=20$ K. $Q_{X}$ also increases with decreasing $T_{0}$, as expected.

\begin{figure*}
\centering
\includegraphics[width=0.7\textwidth]{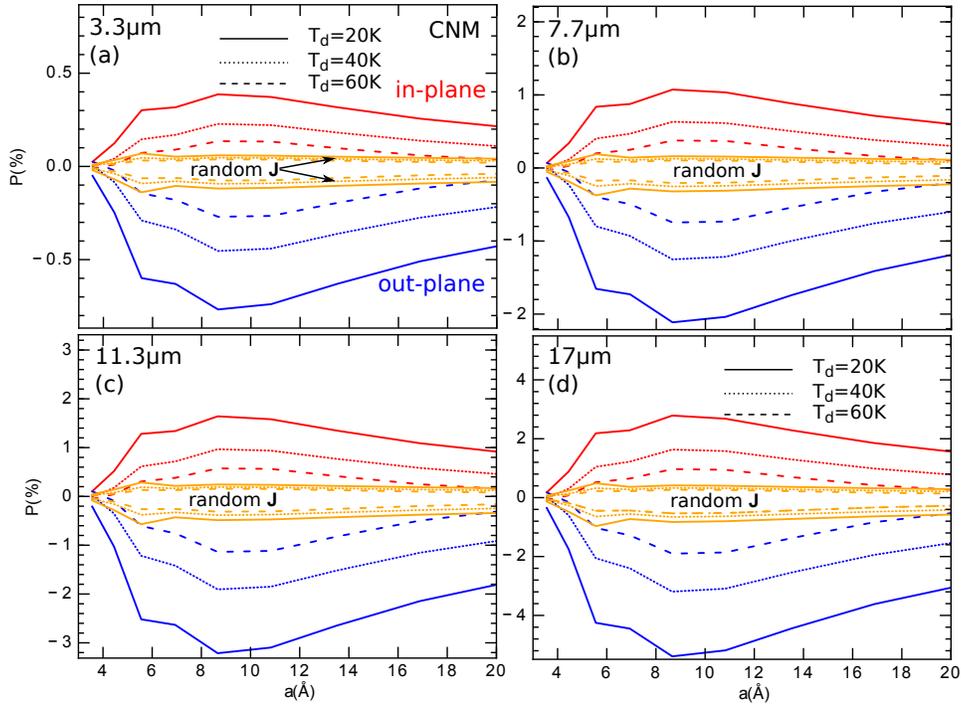}
\caption{The polarization degree of PAH emission features computed for the different values of $T_{0}$. Orange lines show the results for the case of random orientation of $\bJ$ in the space. Perpendicular illumination of $\alpha=\pi/2$ is assumed. The polarization level peaks around $a\sim 8.5$\AA, the same position as $Q_{J}$.}
\label{fig:pol_CNM}
\end{figure*}

Figure \ref{fig:pol_CNM} shows the polarization degree predicted for the strong PAH emission features from CNM. We also show the results computed for the case of the random orientation of the grain angular momentum in the space, which were obtained by plugging our calculated values of $\gamma_{\rm sup,0,ir}$ into Equations (23) and (24) of SD09 (orange lines). The effect of external alignment is negligible for $Q_{J}\le 0.01$ and becomes significant when $Q_{J}$ is maximum at $a \sim 10$\AA. We see that a degree of external alignment $Q_{J}\sim 0.04$ can increase the polarization degree by a factor of 2, e.g., from $-0.5\%$ to $-1.2\%$ for the 11.3$\mum$ (Figure \ref{fig:pol_CNM}(c), lower panel). The polarization level is stronger for the longer emission wavelength due to higher $\gamma_{\rm ir}$ (i.e., higher internal alignment). 
Moreover, the polarization degree is small for the typical temperature of $T_{0}=60$ K, and increases significantly with decreasing $T_{0}$, as expected from the corresponding increase of the degree of alignment. The polarization level can reach above $2\%$ for $T_{0}=20$ K.

\subsection{Reflection Nebulae}
\begin{figure*}
\centering
\includegraphics[width=0.7\textwidth]{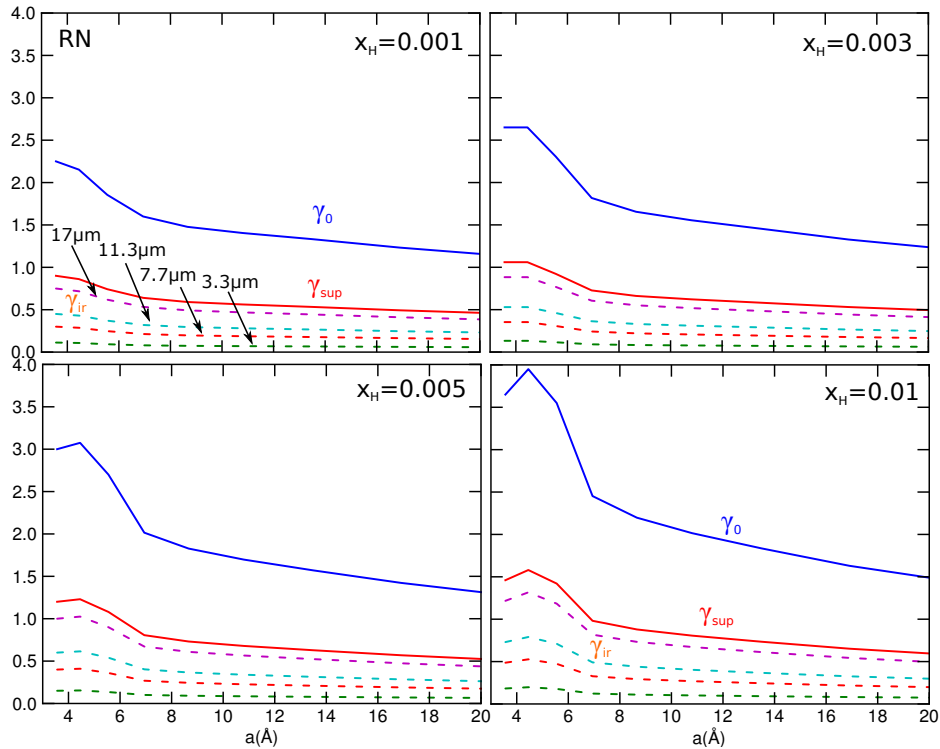}
\caption{The values of $\gamma_{\rm sup},\gamma_{0}$ and $\gamma_{\rm ir}$ computed for the different ionization fractions $x_{\H}$ in the RN. The typical temperature $T_{0}=40$K is considered.}
\label{fig:Trot_Td_RN}
\end{figure*}

Next, we consider the RN conditions in which grain rotational damping and excitation by IR emission is dominant due to intense stellar radiation (\citealt{1998ApJ...508..157D}). Since the ionization fraction $x_{\H}$ is expected to vary across the RN, we consider four different values of $x_{\H}=0.001, 0.003, 0.005$ and $0.01$. The obtained results of $\gamma_{\rm sup}, \gamma_{0}, \gamma_{\rm ir}$ are shown in Figure \ref{fig:Trot_Td_RN}. 

The rotational temperature increases significantly with decreasing the grain size $a$ due to increasing excitation by ion collisions that act on negatively charged small PAHs (HDL10; HL16). This results in the similar trend of $\gamma_{\rm sup}, \gamma_{0}, \gamma_{\rm ir}$, as seen Figure \ref{fig:Trot_Td_RN}. Moreover, these alignment parameters increase with increasing $x_{\H}$, and smallest PAHs can reach suprathermal rotation for $x_{\H}\ge 0.005$.

\begin{figure}
\includegraphics[width=0.45\textwidth]{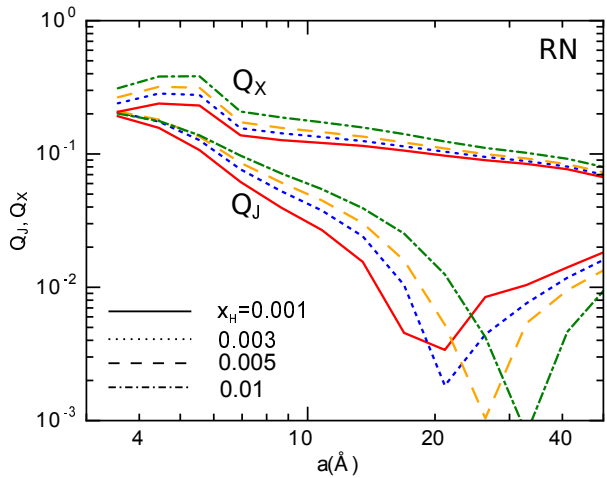}
\caption{Degree of internal alignment ($Q_{X}$) and external alignment ($Q_{J}$) of PAHs in the RN for a variety of $x_{\H}$. The value of $Q_{J}$ rises toward smaller $a$, the same trend as $\gamma_{\rm sup}$.} 
\label{fig:QXQJ_RN}
\end{figure}

Figure \ref{fig:QXQJ_RN} shows $Q_{X}$ and $Q_{J}$ for the different values of $x_{\H}$. Both $Q_{X}$ and $Q_{J}$ increase with increasing $x_{\H}$ due to the faster rotation (same as $\gamma_{\rm sup}$). The external alignment $Q_{J}$ increases with decreasing $a$ and can reach $\sim 0.2$ for the smallest PAHs due to the increase of $\gamma_{\rm sup}$ (see Figure \ref{fig:Trot_Td_RN}).

\begin{figure*}
\centering
\includegraphics[width=0.7\textwidth]{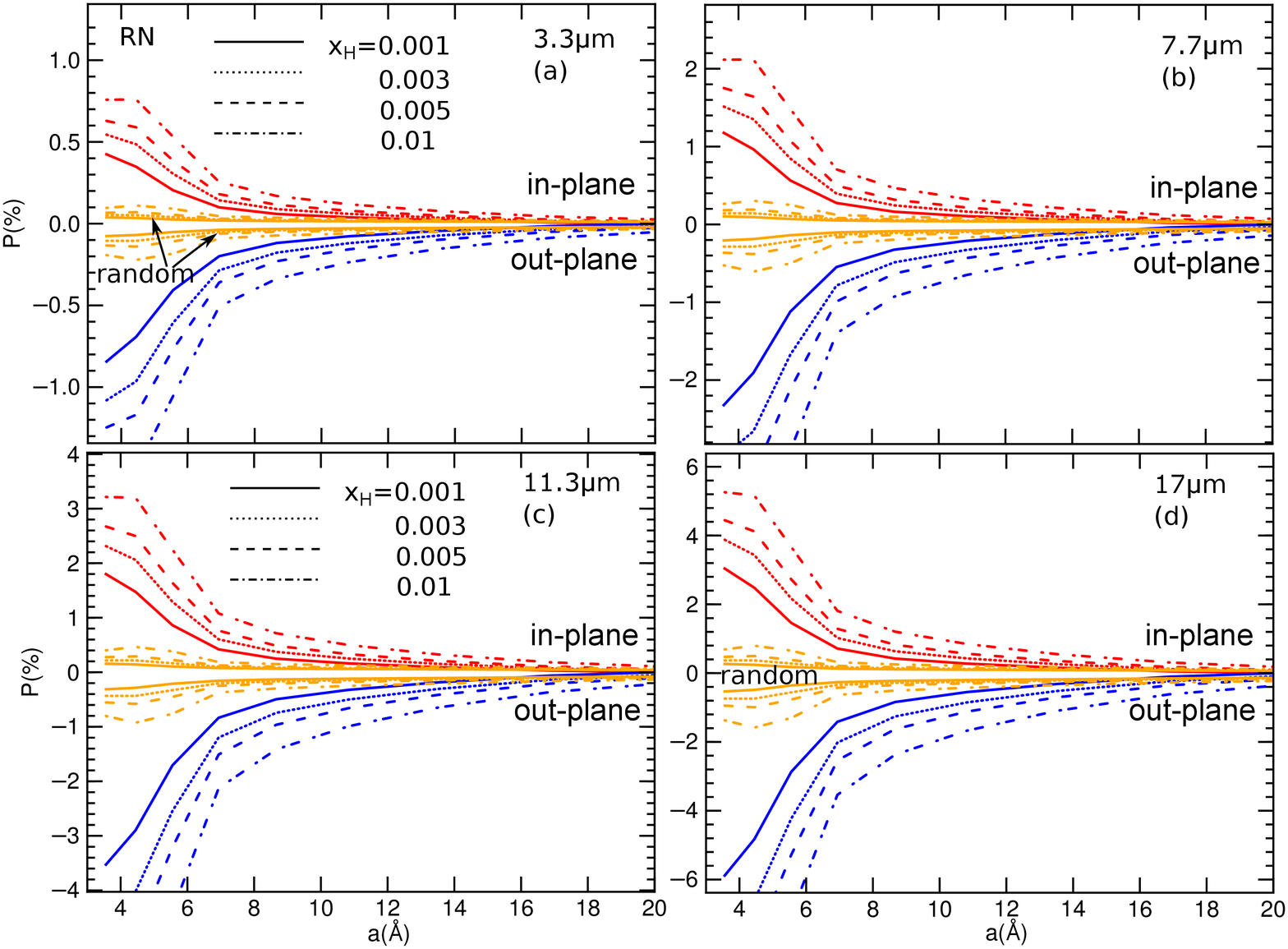}
\caption{Polarization degree of PAH emission features calculated for the RN. Four different levels of ionization $x_{\H}$ are considered. The polarization degree increases toward smaller $a$, the same trend as $Q_{J}$ and $\gamma_{\rm sup}$.}
\label{fig:pol_PAH_RN}
\end{figure*}

Figure \ref{fig:pol_PAH_RN} shows the polarization degree predicted for the different emission features as a function of $a$. The polarization level tends to rapidly increase with decreasing $a$ for $a<10$\AA, whereas the polarization obtained for the case of random angular momentum increases rather slowly. Such an enhancement in the polarization level is produced by the significant increase in $Q_{J}$ that rises beyond $5\%$ (see Figure \ref{fig:QXQJ_RN}). This is because the smaller PAHs can be spun-up to higher rotational temperature as shown in Figure \ref{fig:pol_PAH_RN}. The polarization level increases with the ionization fraction $x_{\H}$, as expected.

\begin{figure*}
\centering
\includegraphics[width=0.7\textwidth]{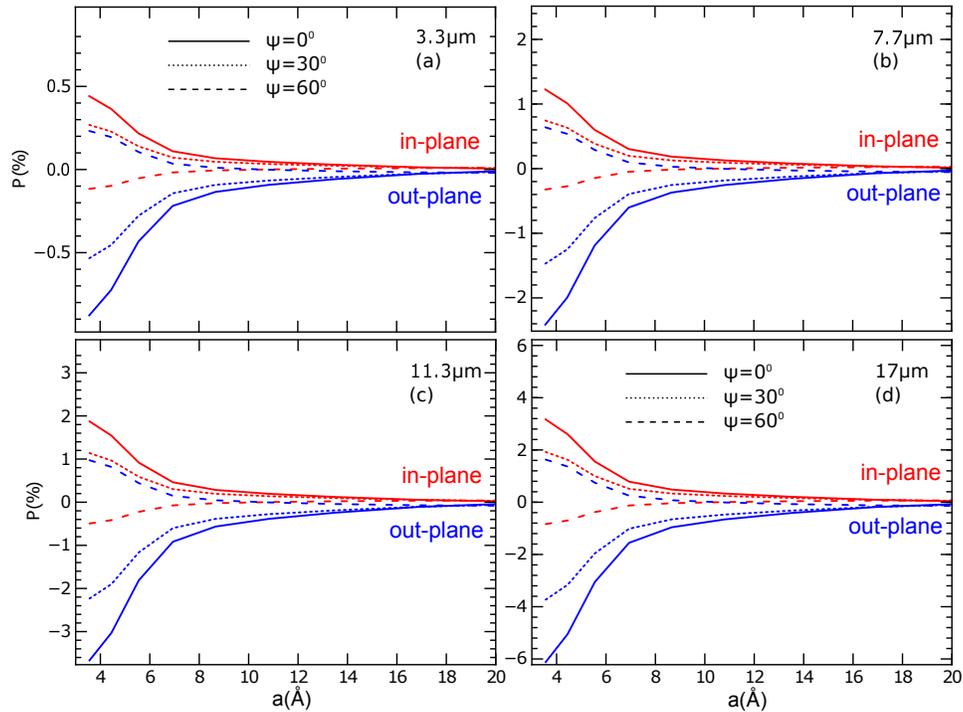}
\caption{Polarization of PAH emission features for the different directions of the magnetic field relative to the plane of the sky with $\psi=0, 30, 60^{\circ}$. The polarization reverses its sign for $\psi=60^{\circ}$.}
\label{fig:pol_PSI_RN}
\end{figure*}

To study the effect of the magnetic field direction, we compute the polarization level for two additional directions with $\psi=30^{\circ}$ and $60^{\circ}$. The results for the case $x_{\H}=0.001$ are shown in Figure \ref{fig:pol_PSI_RN}. The polarization level decreases with increasing $\psi$. In particular, we find that the polarization changes its sign for $\psi=60^{\circ}$.

\subsection{PDR and Orion Bar}
PDR conditions are expected to be the most favorable conditions for detecting polarized PAH emission (SD09). Here, we consider two cases of PDRs, a typical PDR and Orion Bar (see Table \ref{tab:ISM} for their parameters). 

\begin{figure*}
\centering
\includegraphics[width=0.8\textwidth]{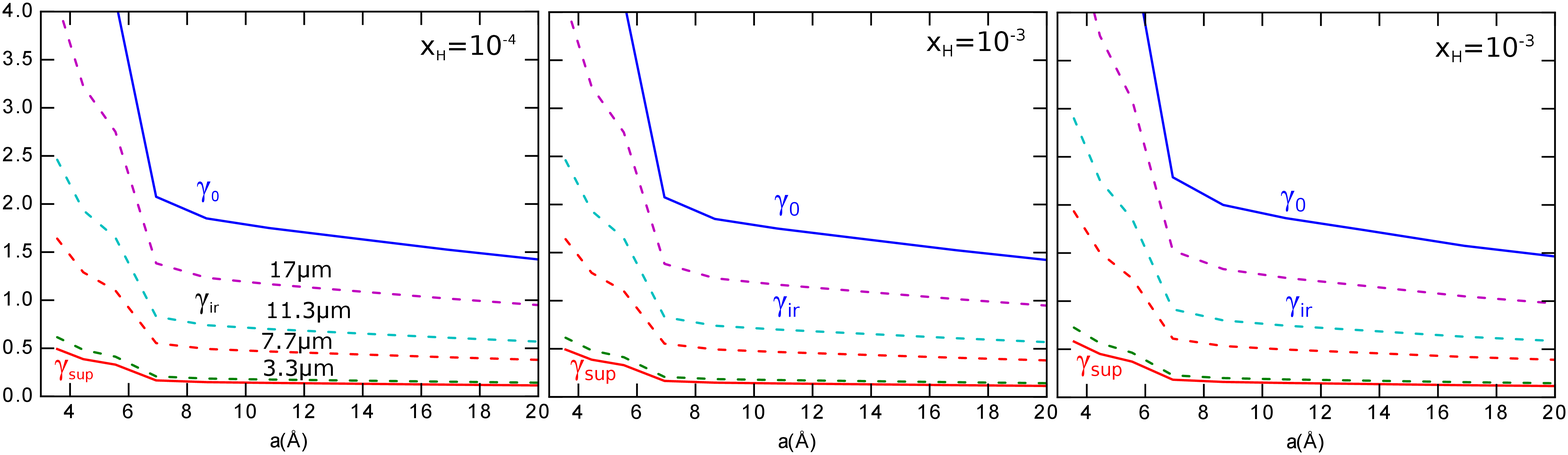}
\caption{Same as Figure \ref{fig:Trot_Td_RN} but for the PDR. Three values $x_{\H}=10^{-4}, 10^{-3}$ and $10^{-2}$ are considered.}
\label{fig:Trot_Td_PDR}
\end{figure*}

Figure \ref{fig:Trot_Td_PDR} shows $\gamma_{\rm sup,0,ir}$ computed for three values of $x_{\H}$ for the PDR.
The rotation of PAHs is mostly subthermal with $\gamma_{\rm sup}\sim 0.2$ for $a> 8$\AA~ and increases with decreased $a$. The rotation slightly increases when the ionization increases from $x_{\H}= 10^{-4}$ to $10^{-3}$. When the ionization is enhanced significantly to $x_{\H}=10^{-2}$, the rotation becomes faster, resulting in a considerable increase of $\gamma_{sup,0,ir}$.

\begin{figure*}
\centering
\includegraphics[width=0.7\textwidth]{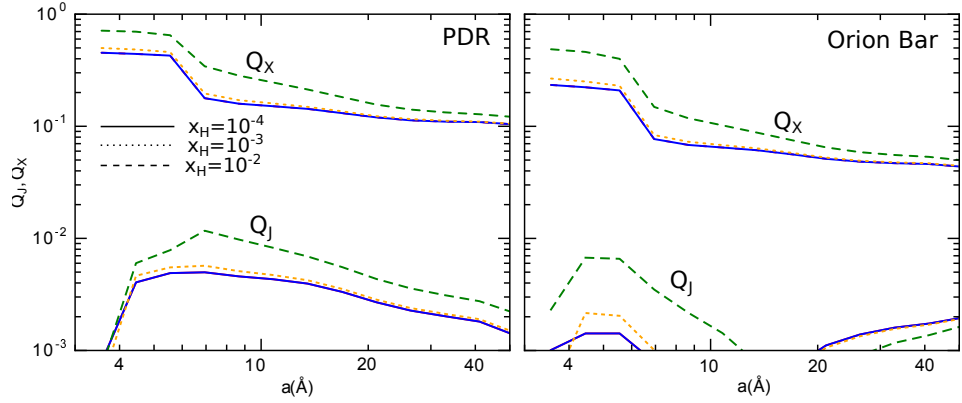}
\caption{Same as Figure \ref{fig:QXQJ_RN} but for the PDR (left) and Orion Bar (right). The internal alignment of the PDR is larger due to lower $T_{0}$.}
\label{fig:QXQJ_PDR}
\end{figure*}

The degrees of grain alignment are shown in Figure \ref{fig:QXQJ_PDR} for PDR (left) and Orion Bar (right), respectively. The degree of external alignment is rather small, with $Q_{J}\le 0.01$. The internal alignment is $Q_{X}\sim 0.1$ for $a>8$\AA~ and increases with decreasing $a$ due to the increase of $\gamma_{\rm sup}$ (Figure \ref{fig:Trot_Td_PDR}). 

\begin{figure*}\centering
\includegraphics[width=0.7\textwidth]{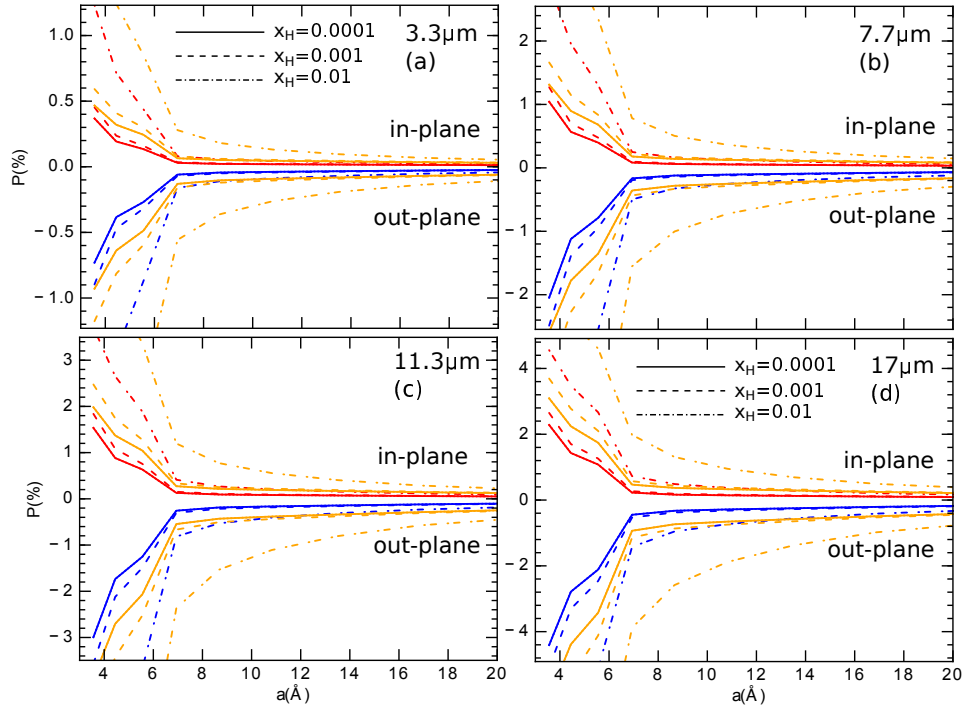}
\caption{Same as Figure \ref{fig:pol_PAH_RN} but for the PDR.}
\label{fig:pol_PDR}
\end{figure*}

The polarization levels for the PDR are shown in Figure \ref{fig:pol_PDR}. The polarization degree is rather small of $P<0.5\%$ for $a>8$\AA~ due to subthermal rotation of $\gamma_{\rm sup}\sim 0.2$ (Figure \ref{fig:pol_PDR}). For smaller PAHs, the rotational temperature increases (Figure \ref{fig:Trot_Td_PDR}), enhancing the polarization level. Since the degree of external alignment is negligible, it does not affect on the polarization because the value of $Q_{J}$ has opposite trend with $P$ for $a<10$\AA~(see Figure ~\ref{fig:QXQJ_PDR}). {Note that our results are slightly lower than the SD09's model. This may arise from the uncorrelation of internal alignment and external alignment (\citealt{1999MNRAS.305..615R}) that is important for negligible external alignment $Q_{J}<0.01$.}

\begin{figure*}\centering
\includegraphics[width=0.7\textwidth]{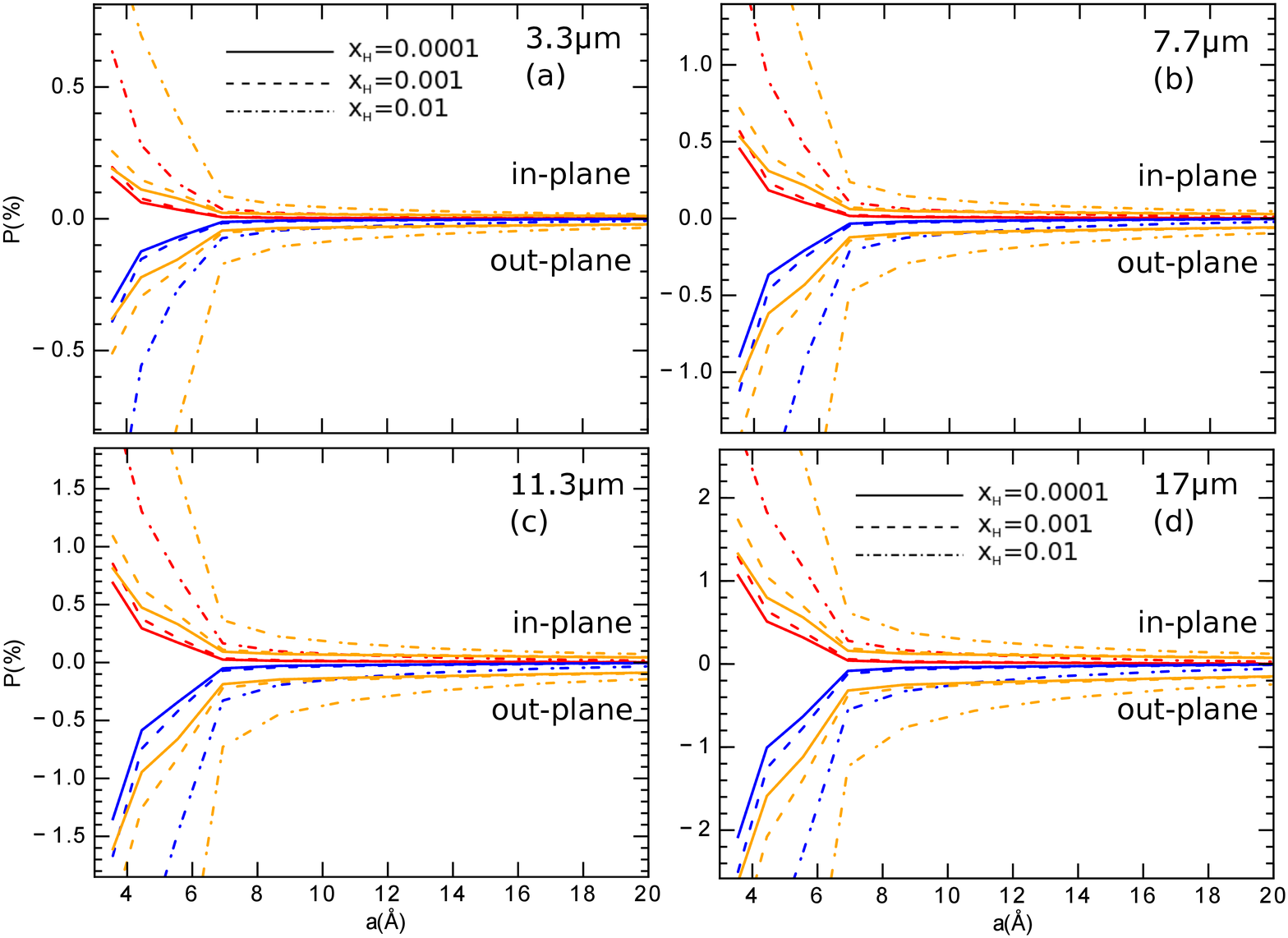}
\caption{Same as Figure \ref{fig:pol_PAH_RN} but for Orion Bar. The size dependence polarization has similar behavior as those in the PDR, but with lower polarization level due to lower degree of grain alignment.}
\label{fig:pol_OrionBar}
\end{figure*}

Figure \ref{fig:pol_OrionBar} shows the results for Orion Bar. Compared to the PDR, the polarization is lower due to higher dust temperature $T_{0}$. As in the RN and PDR, the polarization increases with decreasing size for $a<10$\AA~ and with the ionization fraction $x_{\H}$.

\section{Discussion}\label{sec:discus}

\subsection{Comparison to previous theoretical studies}
\cite{1988prco.book..769L} (L88) presented the first model of polarized PAH emission, where the internal alignment of PAHs with the angular momentum is considered. Assuming random orientation of the angular momentum in the RN conditions, L88 obtained the polarization level of $\sim -2.1\%$ and $\sim0.9\%$ for $11.3\mum$ and $3.3\mum$ features, respectively. SD09 have revisited the L88's model by introducing the parameters for internal alignment before and during the UV absorption and using the realistic dynamics of PAHs. For the similar conditions, SD09 found that the polarization level is only $\sim -0.53\%$ for the $11.3\mu$m feature, and a negligible polarization for $3.3\mum$. In this work, we extended the SD09's model by incorporating the effect of grain alignment due to resonance paramagnetic relaxation. Our model naturally incorporates realistic rotational dynamics of PAHs as in SD09. Moreover, we calculate the polarization for a wide range of grain sizes and environment conditions. 

In the typical conditions of CNM and RN, we found that PAHs rotate at slightly sub-thermal speeds, with suprathermal parameter $\gamma_{\rm sup}\le 1$. The degree of external alignment by resonance paramagnetic relaxation is found to reach several percents. As a result, we found that such external alignment can result in a significant increase in the polarization level of PAH emission compared to the random distribution of $\bJ$. For example, an external alignment of $Q_{J}\sim 0.05$ can enhance the polarization level by several times. For the PDR and Orion Bar, we found that PAHs rotate subthermally, with $\gamma_{\rm sup}<0.5$, which results in a negligible degree of external alignment ($Q_{J}<1\%$) and negligible effect on the polarized PAH emission. For very small PAHs with $a<8$\AA~ that tend to have negative charge (see \citealt{2001ApJS..134..263W}), we found that rotational excitation by ion collisions can enhance $\gamma_{\rm sup}$, resulting in the increase of the polarization level for both the RN and PDRs conditions.

\subsection{Theoretical predictions vs. Observations}

We calculated the polarization of PAH emission from the CNM for several dust temperatures $T_{0}=20, 40$ and 60 K. For the typical temperature $T_{0}=60$ K, the polarization level peaks at $a\sim 9$\AA~ and declines rapidly for smaller PAHs. The peak polarization is less than $0.5\%$ for $3.3\mum$ and $7.7\mum$ (in-plane mode), but increases to $\sim-1\%$ and $-2\%$ for $11.3$ and $17\mum$ (out-of-plane mode). The polarization level is significantly increased to $P\sim 1-5\%$ for $T_{0}=20$ K. 

For the RN conditions, we found that the polarization of PAH emission is significant, which can be in the range from $-1.5\%$ at $a=8$\AA~ to $-4\%$ at $a=4$\AA. For the PDR and Orion Bar, the polarization is rather small, although the smallest ones can emit detectable polarized emission. 

\begin{figure}
\includegraphics[width=0.45\textwidth]{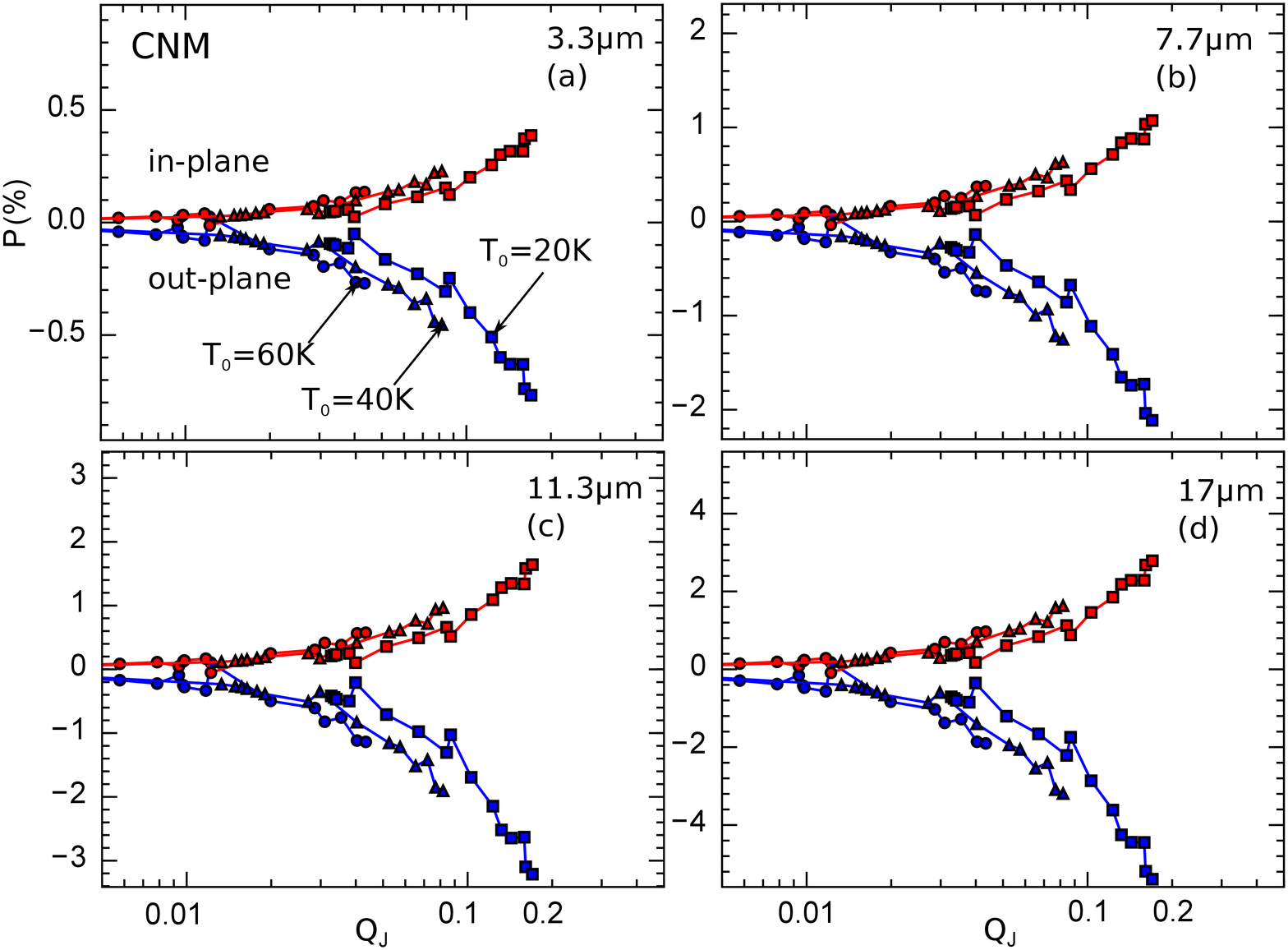}
\caption{Polarization degree vs. $Q_{J}$ for the different grain temperatures from the CNM. }
\label{fig:pol_QJ_CNM}
\end{figure}

\begin{figure}
\includegraphics[width=0.48\textwidth]{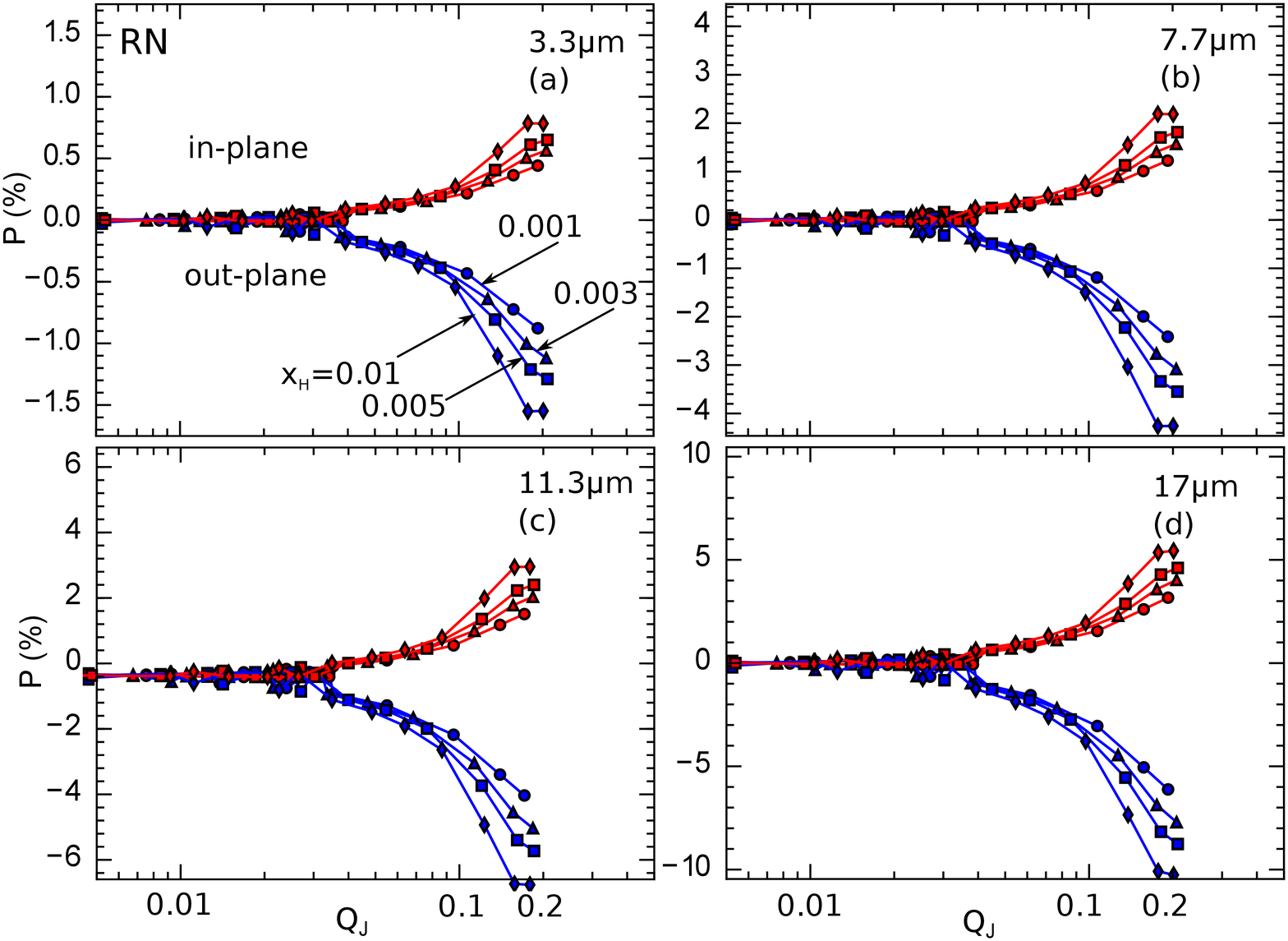}
\caption{Polarization degree vs. $Q_{J}$ for RN with the different values of $x_{\H}$}
\label{fig:pol_QJ_RN}
\end{figure}

Figure \ref{fig:pol_QJ_CNM} and \ref{fig:pol_QJ_RN} show the explicit dependence of the polarization level with the degree of external alignment $Q_{J}$ for CNM and RN, respectively. For both phases, the polarization level is very low for $Q_{J}<0.05$, and it increases rapidly with $Q_{J}$ when the external alignment becomes $Q_{J}\ge 0.05$. 

To date, observational studies of polarized PAH emission are still limited. {Early work by SDL88 reports the detection of polarized PAH emission from the Orion ionization front, with $P\sim 0.86\pm 0.28\%$ for 3.3 $\mu$m at 3.1$\sigma$ limit. Our theoretical results reveal that this high polarization level likely requires a moderate efficiency of external alignment $Q_{J}\sim 0.2$ (Figure \ref{fig:pol_QJ_RN}(a)). This can be fulfilled if emitting PAHs are smaller than 6 \AA~ (see Figure \ref{fig:QXQJ_RN}) for which the rotation rate of PAHs is enhanced due to ion collisions. We note that due to the atmospheric opacity, the ground-based measurement at $3.3\mum$ is very challenging, thus, the detection by SDL88 should be cautious. SDL88 also report a polarization level of $-1.78\pm 0.89\%$ for 11.3$\mum$ at 2$\sigma$ limits, which appears to be a tentative detection only. If future observations report a real detection of polarization at 11.3$\mum$, it would provide convincing evidence for the role of external alignment.}


\subsection{A synergetic approach to constrain alignment mechanism of PAHs and nanoparticles}
In addition to mid-IR emission, spinning PAHs emit microwave rotational emission. This emission process is a leading origin of anomalous microwave emission (AME) in the 10-60 GHz frequency-- a new, important Galactic foreground component, which was discovered about 20 years ago (\citealt{Kogut:1996p5303}; \citealt{Leitch:1997p7359}). The polarization of spinning dust emission poses a challenge for the CMB B-mode detection, and an accurate determination of its polarization level requires a solid understanding of alignment of nanoparticles.

The promising mechanism for alignment of nanoparticles is resonance paramagnetic relaxation (\citealt{2000ApJ...536L..15L}; \citealt{2014ApJ...790....6H}). Recently, \cite{2016ApJ...831...59D} pointed out that quantization effect of vibrational energy within nanoparticles may suppress the energy transfer from the rotational system to the vibrational system, reducing the degree of magnetic alignment. Meanwhile, \cite{2016MNRAS.457.1626P} suggested that {\it Faraday rotation braking} can facilitate the dissipation of rotational energy of diamagnetic grains that enhances internal alignment. A unique way to test the alignment mechanism of nanoparticles is through the polarization of infrared PAH emission. If future observations support our theoretical results obtained using the resonance paramagnetic relaxation, then either quantum suppression is not efficient, or the presence of additional effects (e.g., Faraday braking) that allows the VRE exchange to facilitate the alignment of nanoparticles. Moreover, we demonstrate that the dust temperature before the UV absorption $T_{0}$ is an important parameter that characterizes the degree of external alignment, and the polarization of PAH emission tends to increase with decreasing $T_{0}$. In the light of this result, the non-detection of polarized PAH emission from the diffuse ISM supports the idea that the VRE exchange is frozen at some high temperature.

\subsection{The AME polarization and the exact carrier of the AME}

Previous theoretical predictions using the resonance paramagnetic relaxation mechanism show that the polarization degree of spinning dust emission can be up to a few percents around $10$ GHz and falls to below $1\%$ at $\nu>20$ GHz \citep{2013ApJ...779..152H}. Therefore, this polarized spinning dust emission appears to be a challenge for the detection of CMB B-mode signal, because simulations \citep{2016MNRAS.458.2032R} show that a level of 1$\%$ spinning dust polarization would affect the CMB B-mode detection.

Nevertheless, the exact origin of AME is still debated. In addition to spinning PAHs, spinning nanosilicates \citep{2016ApJ...824...18H}, and nanoiron \citep{2016ApJ...821...91H} are also suggested as the potential carriers of the AME. The AME polarization is thus an important signature to differentiate those carriers. The spinning dust polarization is predicted to be within a few percents (\citealt{2013ApJ...779..152H}; \citealt{2016ApJ...824...18H}) if nanoparticles are aligned by resonance paramagnetic relaxation. If future observations of polarized PAH emission will support our theoretical predictions obtained in this paper, then resonance paramagnetic relaxation alignment would be tested. As a result, the prediction of polarized spinning dust emission from our previous studies (\citealt{2013ApJ...779..152H}; \citealt{2016ApJ...824...18H}) using resonance relaxation would be supported accordingly. As a consequence, spinning iron nanoparticles would be ruled out as a source of the AME due to its high polarization (\citealt{2016ApJ...821...91H}), while spinning PAH, silicate nanoparticles, and magnetic dipole emission remains important candidates for the AME.

\subsection{Studying circumstellar magnetic fields via mid-IR polarization}
The alignment of PAHs with the magnetic field potentially opens a new window in to studying magnetic fields via mid-IR polarization of PAH emission. The polarization direction of in-plane (out-of-plane) mode is perpendicular (parallel) to the magnetic field, while the degree of polarization increases with the strength of the magnetic field. The polarized PAH emission would be most useful for tracing magnetic fields in the environments where strong PAH emission features are observed, such as RN, and circumstellar disks around Herbig Ae/Be stars \citep{2004A&A...427..179H} and T-Tauri stars. 

{We note that the PAH structure and abundance are sensitive to environments (e.g., \citealt{2016A&A...590A..26C}). Small PAHs are expected to be destroyed by photodissociation in intense radiation fields, although some strongest PAHs (namely grandPAH) may still survive in bright PDRs (\citealt{2015ApJ...807...99A}). \cite{2017ApJ...835..291S} have derived the size distribution of PAHs in protoplanetary disks (PPDs) around T-Tauri and Herbig Ae/Be stars. Their modeling of PAH emission from 62 PPDs shows that the peak grain size $a_{p}$ is mostly below $10$\AA~(i.e., only four PPDs with $a_{p}>10$\AA; see Table 2 in \citealt{2017ApJ...835..291S}). Interestingly, small PAHs (up to 70 C atoms) are also seen in NGC 7023 nebula by \cite{2016A&A...590A..26C}. Therefore, the potential to observe polarized PAH emission by small PAHs and to trace magnetic fields in PPDs is promising.}

\section{Summary}\label{sec:sum}
We have calculated the polarization of PAH emission for the different environment conditions by incorporating the alignment of PAHs with the magnetic field, and discussed broad implications of this study. Our principal results are summarized as follows:

\begin{itemize}

\item[1] We computed the polarization of PAH emission using the data of grain angular momentum that is obtained from simulations of PAH alignment by resonance magnetic relaxation for the CNM, RN, PDR, and Orion Bar. We found that a considerable degree of external alignment (i.e., $Q_{J}$ above several percent) can increase the polarization level of PAH emission, compared to the results obtained with randomly oriented $\bJ$. The polarization degree tends to increase with increasing the ionization fraction.

\item[2] Our results for the CNM show that the polarization level is sensitive to the grain temperature, and the polarization can be large if PAHs can cool down to $T_{0}\sim 20$ K. This feature can be useful to probe the vibrational-rotational energy transfer process.

\item[3] We found that the polarization of PAH emission by very small PAHs can be large for the RN conditions in which the external alignment $Q_{J}$ can reach $10\%$, suggesting RN as the most favorable conditions to observe polarized PAH emission. 

\item[4] We obtained a relationship between the polarization level and the degree of external alignment. This can be used to constrain the alignment of PAHs and test alignment theory of nanoparticles through observations, leading to better constraints on the polarization of anomalous microwave emission from rapidly spinning nanoparticles. 

\item[5] For the typical conditions of the PDR and Orion Bar, we found that the external alignment is small ($Q_{J}<0.01$), resulting in a low polarization level. In the regions with enhanced ionization fraction of $x_{\H}\sim 10^{-2}$, the polarization can be increased considerably. 

\item[6] We found the polarization of PAH emission (both direction and level) is a function of the magnetic field direction. This may be used study the magnetic fields in the conditions with known illumination directions.
\end{itemize}

\begin{acknowledgements}
We are very grateful to an anonymous referee for a careful reading and constructive comments that helped us improve the clarity of the paper. We thank B-G Andersson for sending us a reference on graphene magnetism by hydrogen adsorption. We thank Bruce Draine, Alex Lazarian, Peter Martin, and Charles Telesco for insightful discussions. Numerical calculations are conducted by using a high-performance computing cluster at the Korea Astronomy and Space Science Institute (KASI).
\end{acknowledgements}

\appendix
\section{Absorption and Emission Cross-section of PAHs}\label{apd:A}
{L88 provided an excellent description of PAH absorption and emission, assuming that the PAH geometry is described by a thin disk. Here we provide a short summary for reference.}

\subsection{Cross-Section of Absorption}

For an instantaneous orientation of the PAH in which the symmetry axis $\ahat_{1}$ makes an angle $\Theta$ with the radiation direction, the grain cross-section for absorption of unpolarized starlight is given by
\bea
A_{\star}\propto \langle |\bE.{\bf d}|^{2}\rangle
\ena
where ${\bf d}$ is the electric dipole of the PAH. For the absorption of UV photons that are induced by $\pi-\pi*$ electronic transitions, the transition dipoles lie in the PAH plane. Thus, we can denote ${\bf d}=(0, 1, 1)$ along three axes $\ahat_{1}\ahat_{2}\ahat_{3}$.

For an unpolarized radiation source, it is easy to show that
\bea
A_{\star}\propto 1+\cos^{2}\Theta,
\ena
where $\Theta$ is the angle between $\ahat_{1}$ and the radiation direction $\bk$, such that $\cos\Theta=\ahat_{1}.\zhat_{k}$. {Therefore, the cross-section $A_{\star}$ at $\Theta=0$ is two times higher than that at $\Theta=90^{\circ}$. This is easy to understand because at $\Theta=0$, both electric field components of the starlight can excite the oscillation of the PAH dipoles because these electric field components are parallel to the PAH plane (dipoles). In contrast, at $\Theta=90^{\circ}$, only one $E$ component parallel to the PAH plane can be absorbed. The higher cross-section will result in higher IR emission in the two directions, leading to the polarized PAH emission.}


By averaging over the precession of $\ahat_{1}$ around $\bJ$, one derives
\bea
A_{\star}(\beta,\theta)=1+\cos^{2}\beta\cos^{2}\theta + \frac{1}{2}\sin^{2}\beta\sin^{2}\theta=1+C(\beta,\theta),
\ena
where $C(\beta,\theta)$ is the last two terms. In the above equation, $\beta$ describes the orientation of $\bJ$ relative to the anisotropic direction of radiation $\zhat$ (Figure \ref{fig:RF}(b)), and $\theta$ determines the angle of $\ahat_{1}$ with $\bJ$ (Figure \ref{fig:RF}(c)).

For a galactic disk, PAH is illuminated by radiation from all direction. Let the incident direction determined by the angle $\psi'$ and $\theta'$. Assuming the axisymmetric structure, the azimuthal angle can be averaged, and one obtains the cross-section for the incident ray $\theta'$ as follows:
\bea
A_{\rm gal}(\beta,\theta,\theta')=1+\cos^{2}\theta' C(\beta,\theta) + \sin^{2}\theta' S(\beta,\theta),
\ena
where
\bea
S(\beta,\theta) = \frac{1}{4}(1+\cos^{2}\beta)\sin^{2}\theta + \frac{1}{2}\sin^{2}\beta\cos^{2}\theta
\ena

\subsection{Cross-section of PAH emission}
The radiation intensity of PAH emission with the electric field along a direction $w$ is given by
\bea
F_{w}\propto \langle |{\bf w}.{\bf d}|^{2}\rangle,
\ena
where ${\bf d} =\sum_{i}\alpha_{i}\ahat_{i}$ is the electric dipole with polarizability $\alpha_{i}$ along the principal axis $\ahat_{i}$. For the in-plane stretching modes, the non-vanishing dipole matrix corresponds to the dipole moments in the PAH plane only, i.e., ${\bf d}/|d|=(0,1,1)$. In contrast, for out-of-plane bending modes, the non-vanishing dipole corresponds to the dipole moment perpendicular to the PAH plane, i.e., ${\bf d}/|d|=(1,0,0)$. 

Therefore, for isotropic polarizability, we have 
\bea
F_{w}^{\perp}\propto \langle |{\bf w}.{\ahat_{1}}|^{2}\rangle,~~F_{w}^{\|}=\langle |{\bf w}.({\ahat_{2}+\ahat_{3})}|^{2}\rangle=1-F_{w}^{\perp}
\ena

Emission of IR photons following UV absorption is due to the oscillation of electric dipoles induced by the incident light. The fluxes of radiation with $\bE$ along the $u-$ and $v-$ directions in the plane of the sky (see Figure \ref{fig:RF}(a)) are
\bea
{F}_{u}^{\|}(\beta,\varphi,\theta,\alpha)=1-{F}_{u}^{\perp}(\beta,\varphi,\theta,\alpha),\label{eq:Fupar}\\
{F}_{v}^{\|}(\beta,\varphi,\theta,\alpha)=1-{F}_{v}^{\perp}(\beta,\varphi,\theta,\alpha),\label{eq:Fvpar}
\ena
where ${F}_{w}^{\perp}$ are easily obtained by transformation of coordinate systems:
\bea
F_{u}^{\perp}&=&\frac{1}{2}\left(\cos^{2}\varphi+\cos^{2}\beta\sin^{2}\varphi\right)\sin^{2}\theta+\sin^{2}\beta\sin^{2}\varphi\cos^{2}\beta,\\
F_{v}^{\perp}&=&\cos^{2}\alpha\left[\frac{1}{2}\left(\sin^{2}\varphi+\cos^{2}\beta\cos^{2}\varphi\right)\sin^{2}\theta\right]+\cos^{2}\alpha
\sin^{2}\beta\cos^{2}\varphi\cos^{2}\theta+\sin^{2}\alpha C(\theta,\beta)\nonumber\\
&&+\frac{1}{4}\sin 2\alpha\sin 2\beta \cos\varphi(1-3\cos^{2}\theta).
\ena

In the case of random orientation of $\bJ$, averaging over $\beta$ and $\varphi$ yields the following (see also L88; SD09):
\bea
\bar{F}_{u}^{\perp}(\beta,\theta,\alpha) &=& S(\beta,\theta),\\
\bar{F}_{v}^{\perp}(\beta,\theta,\alpha) &=& \cos^{2}\alpha S(\beta,\theta)+ \sin^{2}\alpha C(\beta,\theta).
\ena

\section{A simple model of polarized PAH emission}\label{apdx:model}
\begin{figure}
\centering
\includegraphics[width=0.7\textwidth]{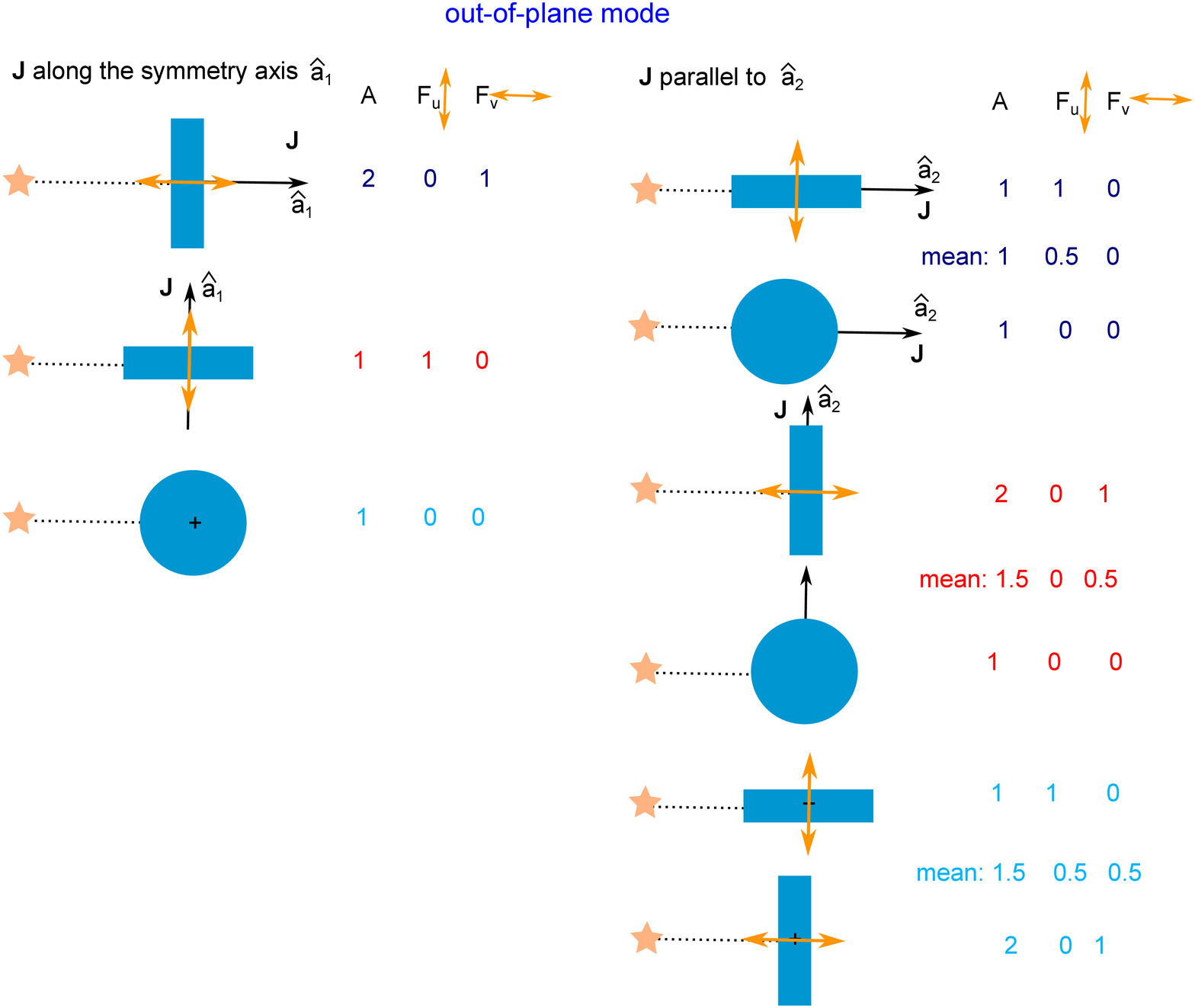}
\caption{Schematic of the polarization of PAH emission by absorption of starlight for the out-of-plane emission mode. Left: perfect internal alignment with $\ahat_{1}\| \bJ$. Right panel: perfect wrong internal alignment with $\ahat_{1}\perp \bJ$.  The polarization is in the illumination direction.}
\label{fig:pol_oop}
\end{figure}

\begin{figure}
\centering
\includegraphics[width=0.7\textwidth]{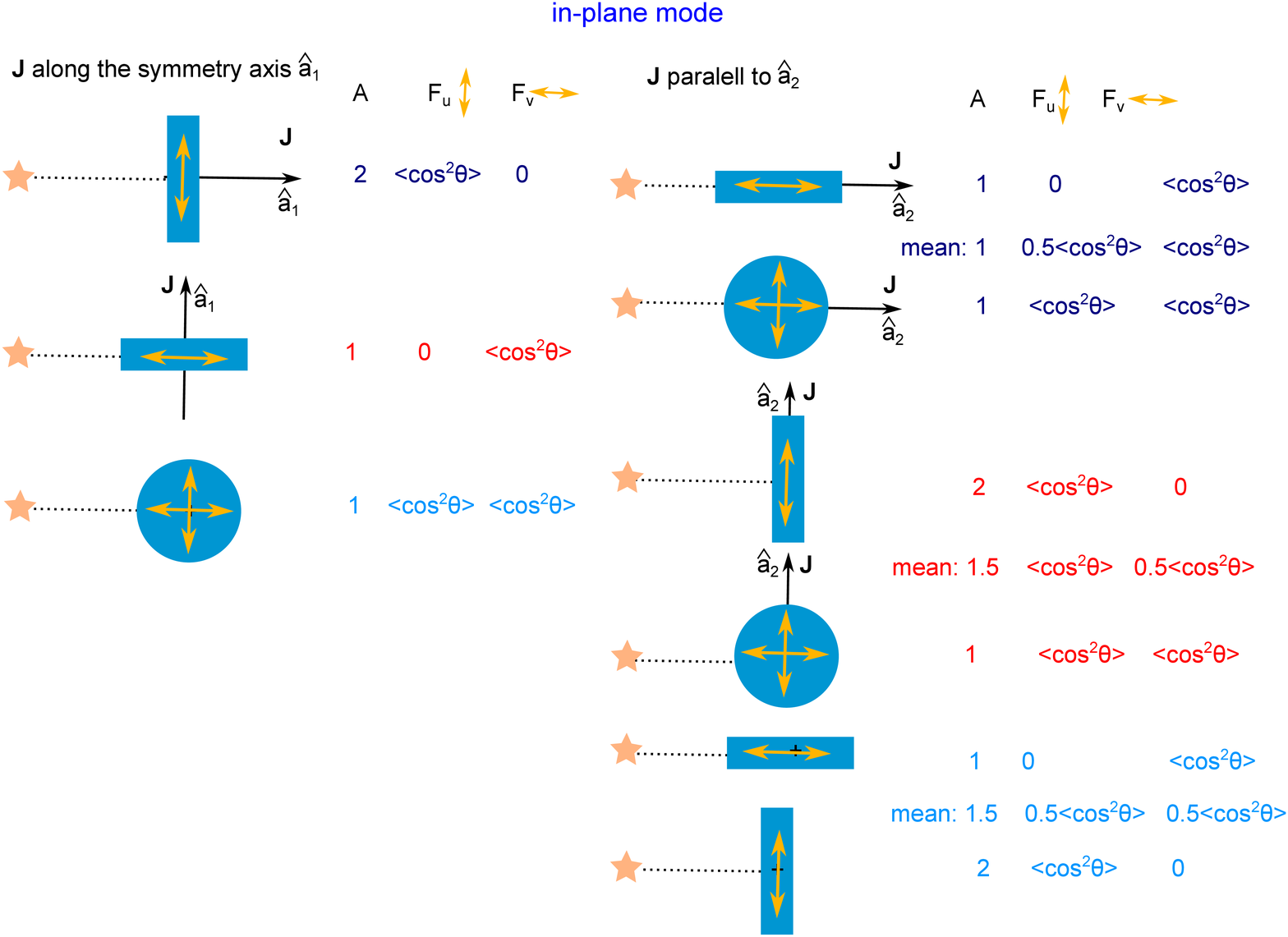}
\caption{Same as Figure \ref{fig:pol_inp} but for the in-plane emission mode. The polarization is perpendicular to the illumination direction. The figure is reproduced from L88.}
\label{fig:pol_inp}
\end{figure}

{To have an intuitive understanding of the mechanism of polarized PAH emission, first let us consider the emission of out-of-plane CH bending mode. Figure \ref{fig:pol_oop} shows the schematic of PAH absorption and emission for two types of internal alignment and three orthogonal orientations of $\bJ$ in the space. The PAH is assumed to be a disk-like shape. The electric dipoles (marked by orange arrows) are radial and follow a uniform distribution.

For $\bJ\|\ahat_{1}$ (perfect internal alignment), using the cross-section $A$ and emission flux coefficients $F_{u}, F_{v}$ from Figure \ref{fig:pol_oop}), we evaluate the mean emission intensity along $\hat{\bf u}-$ and $\hat{\bf v}-$ directions by averaging over three orthogonal orientations of $\bJ$:
\bea
I_{u,\|}=\sum_{i=1}^{3} W_{i}A_{i}F_{u}=\frac{1}{3},\\
I_{v,\|}=\sum_{i=1}^{3} W_{i}A_{i}F_{v}=\frac{1}{3}2,
\ena
where the weight $W_{i}=1/3$ for the isotropic orientation of $\bJ$, and the arbitrary units have been used for the sake of simplicity. The polarization degree is $p=(I_{u,\|}-I_{v,\|})/(I_{u,\|}+I_{v,\|})=-1/3$. We note that the out-of-plane oscillation of the dipole produces the electric field in the sky of the plane, either along $\hat{\bf u}$ or $\hat{\bf v}$.

For $\bJ\perp \ahat_{1}$ (i.e., $\bJ\| \ahat_{2}$ or $\bJ\| \ahat_{3}$), the emission intensity along $\hat{\bf u}-$ and $\hat{\bf v}-$ directions is
\bea
I_{u,\perp}=\sum_{i=1}^{3} W_{i}\langle A_{i}\rangle \langle F_{u}\rangle=\frac{1}{3}(1/2+3/4)=\frac{5}{12},\\
I_{v,\perp}=\sum_{i=1}^{3} W_{i}\langle A_{i}\rangle \langle F_{v}\rangle=\frac{1}{3}(3/4+3/4)=\frac{6}{12},
\ena
where the mean $\langle A\rangle$ is the average over two possible orientations of the grain axis $\ahat_{2}$ with $\bJ$. The polarization becomes $p=(I_{u,\perp}-I_{v,\perp}/(I_{u,\perp}+I_{v,\perp})=-1/11$.

To account for the thermal fluctuations that deviate $\ahat_{1}$ from $\bJ$, we have to average over the internal fluctuations. The weighted emission intensity is 
\bea
I_{u}=\sqrt{2}I_{u,\|} + 2I_{u,\perp}=(\sqrt{2}/3+ 2\times5/12),~I_{v}=\sqrt{2}I_{v,\|} + 2I_{v,\perp}=(2\sqrt{2}/3+2\times6/12),
\ena
where the weight factor $W_{1}=\sqrt{2}$ for the thin disk of inertia moment ratio $h_{a}=2$, and the factor $2$ denotes the case of $\bJ\|\ahat_{2}$ and $\bJ\|\ahat_{3}$ with equal weights.

The polarization level of the in-plane mode is then
\bea
p=\frac{I_{u}-I_{v}}{I_{u}+I_{v}}= -0.196.
\ena

Similarly, Figure \ref{fig:pol_inp} illustrates the polarization of {\it in-plane modes}. We note that for the in-plane mode, the projection of a given radial dipole in the plane of the sky is $\cos\theta$, giving rise to the averaged emission flux of $\langle \cos^{2}\theta\rangle$. 

For $\bJ\|\ahat_{1}$, the emission intensity along $\hat{\bf u}-$ and $\hat{\bf v}-$  directions is given by
\bea
I_{u,\|}=\sum_{i=1}^{3} W_{i}A_{i}F_{u}=\frac{1}{3}3\langle \cos^{2}\theta\rangle=\frac{1}{2},\\
I_{v,\|}=\sum_{i=1}^{3} W_{i}A_{i}F_{v}=\frac{1}{3}2\langle \cos^{2}\theta\rangle=\frac{1}{3},
\ena
where $\langle \cos^{2}\rangle=1/2$ for the uniform distribution of the dipoles in the disk plane. The polarization become $p=(I_{u,\|}-I_{v,\|})/(I_{u,\|}+I_{v,\|})=1/5$.

For $\bJ\|\ahat_{2}$ (or $\ahat_{3}$), the emission intensity along $\hat{\bf u}-$ and $\hat{\bf v}-$  directions is given by
\bea
I_{u,\perp}=\sum_{i=1}^{3} W_{i}\langle A_{i}\rangle \langle F_{u}\rangle=\frac{1}{3}(2+3/4)\langle \cos^{2}\theta\rangle=\frac{11}{24},\\
I_{v,\perp}=\sum_{i=1}^{3} W_{i}\langle A_{i}\rangle \langle F_{v}\rangle=\frac{1}{3}(1+3/2)\langle \cos^{2}\theta\rangle=\frac{10}{24},
\ena
where the mean $\langle A\rangle$ is the average over two possible orientations of the grain axis $\ahat_{2}$ with $\bJ$. The polarization becomes $p=1/21$.

The weighted emission intensity is 
\bea
I_{u}=\sqrt{2}/2+ 2\times11/24, ~I_{v}=\sqrt{2}/3+2\times10/24,
\ena
which leads to the polarization level of $p= 0.12$ (see also L88).

We note that the absorption cross-section $A$ is the same for in-plane and out-of-plane modes because the absorption of UV photons is induced by the transition dipoles in the PAH plane (i.e., $\pi-\pi*$ transitions).}

\section{Coordinate systems and transformation}\label{apd:transform}
The alignment of grain angular momentum $\bJ$ with respect to the magnetic field $\Bv$ is determined by the angles $\xi,\phi$ (see Figure \ref{fig:RF} (d)). The transformation of the basic unit vectors in the $\bJ$-frame to $\Bv$-frame is carried out as follows:
\bea
\zhat_{J}&=&\cos\xi \zhat_{B} + \sin\xi(\cos\phi \xhat_{B}+\sin\phi \yhat_{B}),\\
\xhat_{J}&=&-\sin\xi \zhat_{B} + \cos\xi(\cos\phi \xhat_{B}+\sin\phi \yhat_{B}),\\
\yhat_{J}&=&[\zhat_{J}\times \xhat_{J}]=\cos\phi \yhat_{B}-\sin\phi \xhat_{B}.
\ena

The direction of $\Bv$ in the frame defined by the illumination direction, namely $\kv-$ frame, is determined by the angles $\psi$ and $\zeta$. The transformation of the basic unit vectors from $\Bv$-frame to $\kv$-frame is given by
\bea
\zhat_{B}&=&\cos\psi \zhat_{k} + \sin\psi(\cos\zeta \xhat_{k}+\sin\zeta \yhat_{k}),\\
\xhat_{B}&=&-\sin\psi \zhat_{k} + \cos\psi(\cos\zeta \xhat_{k}+\sin\zeta \yhat_{k}),\\
\yhat_{B}&=&[\zhat_{B}\times \xhat_{B}]=\cos\zeta \yhat_{k}-\sin\zeta \xhat_{k}.
\ena

Finally, the transformation from the $\kv$-frame to the observer's frame follows
\bea
\zhat_{k}=\cos\alpha \zhat_{\rm obs} +\sin\alpha \xhat_{\rm obs},~
\xhat_{k}=-\sin\alpha \zhat +\cos\alpha \xhat_{\rm obs},~
\yhat_{k}=\yhat_{\rm obs},
\ena
where $\xhat_{\rm obs}=\hat{v}, \yhat_{\rm obs}=\hat{u}$ (see Figure 1f).

The transformation between the coordinate system defined by basic unit vectors $\lbrace\ehat\rbrace_{i}$ to a new basis $\lbrace\ehat'\rbrace_{j}$ is done through the matrix transformation:
\bea
\ehat'_{i} = A_{ij}\ehat_{j},
\ena
where $A$ is the matrix of coordinate transformation, which come directly from the above equations.

It follows that 
\bea
\cos\beta &=& \zhat_{J}.\zhat_{k}=\cos\xi\cos\psi -\sin\xi\cos\phi\sin\psi,\\
\sin\beta\cos\varphi &=& \zhat_{J}.\xhat_{k}=\cos\xi\sin\psi\cos\zeta + \sin\xi\left(\cos\phi\cos\psi\cos\zeta -\sin\phi\sin\zeta \right)
\ena

Thus, for a given magnetic field geometry ($\zeta, \psi$), using the simulated data of $\xi, \phi$, we can derive the angles $\beta, \varphi$ to be used to compute the polarization of PAH.

Let $\ahat_{1},\ahat_{2},\ahat_{3}$ be the grain principal axes where $\ahat_{1}$ is the axis of maximum moment of inertia. With the angles defined in Figure \ref{fig:RF}, we have the following:
\bea
\ahat_{1}&=&\cos\theta \zhat_{J} + \sin\theta(\cos\alpha \xhat_{J}+\sin\alpha \yhat_{J}),\\
\ahat_{2}&=&-\sin\theta \zhat_{J} + \cos\theta(\cos\alpha \xhat_{J}+\sin\alpha \yhat_{J}),\\
\ahat_{3}&=&[\ahat_{1}\times \ahat_{2}]=\cos\alpha \yhat_{J}-\sin\alpha \xhat_{J}.
\ena

\section{Degrees of Grain Alignment}\label{apdx:alignment}
Let $Q_{X}=\langle G_{X}\rangle $ with $G_{X}=\left(3\cos^{2}\theta-1\right)/2$ be the degree of internal alignment of the grain symmetry axis $\ahat_{1}$ with $\bJ$, and let $Q_{J}=\langle G_{J}\rangle$ with $G_{J}=\left(3\cos^{2}\xi-1\right)/2$ be the degree of external alignment of $\bJ$ with $\Bv$. Here the angle brackets denote the average over the ensemble of grains. The net degree of alignment of $\ahat_{1}$ with $\Bv$, namely the Rayleigh reduction factor, is defined as $R = \langle G_{X}G_{J}\rangle$ \citep{Greenberg:1968p6020}.

The angular momentum $\bJ$ and the angle $\xi$ obtained from the Langevin equations are employed to compute the degrees of alignment, $Q_{J}, Q_{X}$ and $R$ as follows: 
\bea
 Q_{J}\equiv \sum_{n=1}^{N} \frac{G_{J}(\cos^{2}\xi_{n})}{N}, Q_{X} =\frac{1}{N}\sum_{n=1}^{N}\int_{0}^{\pi}G_{X}(\cos^{2}\theta)f_{\rm LTE}(\theta, J_{n})d\theta.
 \ena
  
\bibliographystyle{apj} 
\bibliography{ms}

\end{document}